\documentclass[aps,prd,10pt,nofootinbib,showpacs,showkeys,superscriptaddress,preprintnumbers,altaffilletter,floatfix,twocolumn]{revtex4-2}

\usepackage{amsmath}
\usepackage{multirow}
\usepackage{hyperref}
\usepackage{soul}
\usepackage[dvipsnames]{xcolor}     % use if color is used in text
\usepackage{graphicx}
\usepackage{dcolumn}
\usepackage{amssymb}
\usepackage{amsfonts}
\usepackage{amsbsy}
\usepackage{booktabs}
\usepackage{color}
\usepackage{rotating}
\usepackage{overpic}
\usepackage{ifxetex}
\usepackage{lipsum}
\usepackage{float}
\usepackage{bbold}
\usepackage{enumerate}
\usepackage[normalem]{ulem}
\usepackage{dsfont}
\usepackage{mathrsfs}
\usepackage{mathtools}
\usepackage{tensor}
\usepackage{caption}
\usepackage{subcaption}
\usepackage{ragged2e}
\ifxetex
   \usepackage{polyglossia}
   \setdefaultlanguage{english}
   \usepackage{fontspec}
\else
   \usepackage[english]{babel}
   \usepackage[utf8]{inputenc}
\fi

\definecolor{amaranth}{rgb}{0.9, 0.17, 0.31}
\definecolor{palatinateblue}{rgb}{0.15, 0.23, 0.89}
\definecolor{brightpink}{rgb}{1.0, 0.0, 0.5}
\hypersetup{ linktoc=all,
    colorlinks, linkcolor={palatinateblue},
    citecolor={brightpink}, urlcolor={amaranth}
}

\graphicspath{ {images/} }

\newcommand{\COMMariam}[1]{}  % Cross Out for commented text

\usepackage{enumitem}

\begin{document}

\title{New Observational Insights into Generalised Axion-Like Dark Energy}
%\title{Dark energy driven by an oscillating generalised axion-like quintessence field}

\author{Mariam Bouhmadi-López}
\email{mariam.bouhmadi[at]ehu.eus}
\affiliation{IKERBASQUE, Basque Foundation for Science, 48011, Bilbao, Spain}
\affiliation{Department of Physics, University of the Basque Country UPV/EHU, P.O. Box 644, 48080 Bilbao, Spain}
\affiliation{EHU Quantum Center, University of the Basque Country UPV/EHU, P.O. Box 644, 48080 Bilbao, Spain}
\author{Hsu-Wen Chiang}
\email{jiangxw[at]sustech.edu.cn, b98202036[at]ntu.edu.tw}
\affiliation{Department of Physics, Southern University of Science and Technology, Shenzhen 518055, China}
\author{Carlos G. Boiza}
\email{c.garcia-boiza[at]ip2i.in2p3.fr}
\affiliation{Institut de Physique des Deux Infinis de Lyon, Université de Lyon, UCBL, UMR 5822, CNRS/IN2P3,
4 rue Enrico Fermi, 69622 Villeurbanne Cedex, France}
\affiliation{Department of Physics, University of the Basque Country UPV/EHU, P.O. Box 644, 48080 Bilbao, Spain}
\affiliation{EHU Quantum Center, University of the Basque Country UPV/EHU, P.O. Box 644, 48080 Bilbao, Spain}
\author{Thomas J. Broadhust}
\email{tom.j.broadhurst[at]gmail.com}
\affiliation{IKERBASQUE, Basque Foundation for Science, 48011, Bilbao, Spain}
\affiliation{Department of Physics, University of the Basque Country UPV/EHU, P.O. Box 644, 48080 Bilbao, Spain}
\affiliation{EHU Quantum Center, University of the Basque Country UPV/EHU, P.O. Box 644, 48080 Bilbao, Spain}

\begin{abstract}
We consider a quintessence model based on a generalised axion-like scalar field potential, originally introduced as a phenomenological framework to describe the late-time accelerated expansion of the Universe. The model exhibits a tracking regime that alleviates the coincidence problem and naturally leads to a phase of late-time acceleration driven by an effective cosmological constant, without requiring fine-tuning of the scalar field initial conditions. In this work, we confront the background cosmological evolution of
this model with current observations by performing a parameter
inference analysis using cosmic microwave background distance priors,
baryon acoustic oscillation measurements, Type~Ia supernovae,
the SH0ES distance ladder, cosmic chronometers, and gamma-ray bursts.
This analysis allows us to assess the observational viability of the
model, its ability to reconcile different cosmological datasets,
and its statistical performance relative to $\Lambda$CDM. %In this work, we confront the background cosmological evolution of this model with current observations by performing a parameter inference analysis using cosmic microwave background shift parameters, baryon acoustic oscillation data, and the SH0ES determination of the local expansion rate. This analysis allows us to assess the consistency of the model with the observed late-time expansion history of the Universe.
\end{abstract}

\maketitle

\renewcommand{\tocname}{Index}

%\tableofcontents

\section{Introduction}\label{intro}

Observations over the past two decades have established that the Universe is currently undergoing a phase of accelerated expansion. This behaviour was first inferred from measurements of luminosity distances to high-redshift Type~Ia supernovae \cite{SupernovaSearchTeam:1998fmf,SupernovaCosmologyProject:1998vns} and has since been independently confirmed by observations of the cosmic microwave background and large-scale structure \cite{Planck2018,Alam2021}. Within the standard $\Lambda$CDM cosmological model, late-time acceleration is attributed to a cosmological constant. While this framework provides an excellent fit to observational data when individual cosmological probes are analysed separately, mild but persistent inconsistencies arise when multiple data sets are combined \cite{CosmoVerseNetwork:2025alb}. Moreover, the physical origin of the cosmological constant remains poorly understood, giving rise to well-known theoretical challenges. Together, these observational and conceptual issues motivate the exploration of extensions of $\Lambda$CDM that go beyond a strictly constant dark energy component \cite{CANTATA:2021asi}.

A broad class of such extensions assumes that dark energy evolves dynamically over cosmic time. Well-studied examples include scalar field models such as quintessence \cite{Ratra:1987rm}, k-essence theories with non-canonical kinetic terms \cite{Armendariz-Picon:2000ulo}, and axion-like dark energy scenarios \cite{Kamionkowski:2014zda,Emami:2016mrt,Chiang:2025qxg,Bouhmadi-Lopez:2025wxo}. More general scalar-tensor constructions, including kinetic gravity braiding and subclasses of Horndeski theories, further enlarge the space of viable dark energy dynamics while preserving second-order field equations \cite{Deffayet:2010qz,Pujolas:2011he,BorislavovVasilev:2022gpp,BorislavovVasilev:2024loq,Kobayashi:2019hrl}. The late-time acceleration of the universe can as well be described by a 3-form \cite{Koivisto:2009fb,Morais:2016bev,Bouhmadi-Lopez:2016dzw} and can help to alleviate the Hubble tension \cite{Bouhmadi-Lopez:2025lzm,Bouhmadi-Lopez:2026ytj}.
Alternative phenomenological approaches describe dark energy as an effective perfect fluid with a parametrised equation of state, enabling systematic studies of departures from $\Lambda$CDM at both the background and perturbative levels \cite{Kamenshchik:2001cp,Bouhmadi-Lopez:2004mpi,Bouhmadi-Lopez:2006fwq}. Closely related scenarios include sign-switching cosmological constant models, in which the effective dark energy density changes sign during cosmic evolution \cite{Akarsu:2021fol,Akarsu:2022typ,Akarsu:2023mfb,Bouhmadi-Lopez:2025ggl,Bouhmadi-Lopez:2025spo,Ghafari:2025eql,Akarsu:2025nns,Bouhmadi-Lopez:2026vyc}.

Alternatively, late-time cosmic acceleration may be interpreted as a consequence of modified gravitational dynamics rather than the presence of a new dark energy component. In this perspective, the accelerated expansion arises from extensions or generalisations of General Relativity, potentially alleviating some of the conceptual difficulties associated with a cosmological constant \cite{CANTATA:2021asi}. Representative examples include $f(\mathcal{R})$ gravity \cite{Capozziello:2011et,Nojiri:2010wj,Nojiri:2017ncd,Morais:2015ooa}, teleparallel extensions such as $f(\mathcal{T})$ gravity \cite{Bengochea:2008gz,Ferraro:2006jd,Cai:2015emx}, and theories based on non-metricity, including $f(\mathcal{Q})$ gravity \cite{BeltranJimenez:2018vdo,BeltranJimenez:2019tme,Ayuso:2020dcu,Boiza:2025xpn,Ayuso:2025vkc}. Collectively, these frameworks provide a broad theoretical laboratory for probing physics beyond $\Lambda$CDM and for confronting the late-time acceleration of the Universe with current and forthcoming observational data.

In this paper, we focus on a class of quintessence models inspired by axion-like dark energy scenarios, motivated by the framework introduced in \cite{Boiza:2024azh,Boiza:2024fmr}. Axion-like particles arise naturally in theories involving spontaneously broken global symmetries and are characterised by periodic scalar field potentials \cite{Marsh:2015xka}. Depending on their characteristic mass and energy scales, these fields can play different roles throughout cosmic history, including driving inflation \cite{Freese:1990rb,Adams:1992bn,Dimopoulos:2005ac,Easther:2005zr,Bachlechner:2014hsa}, contributing to early dark energy \cite{Poulin:2018cxd,Kamionkowski:2022pkx}, or acting as ultralight dark matter candidates \cite{Hu:2000ke,Hui:2016ltb}. In particular, the wave nature of ultralight axion-like dark matter can give rise to distinctive structures on galactic scales, including interference patterns and solitonic cores, as explored in~\cite{Schive:2014dra,Schive:2014hza,Mocz:2019uyd}. At sufficiently small mass scales, axion-like fields may instead provide a dynamical explanation for the late-time accelerated expansion of the Universe \cite{Katewongveerachart:2026ovj,Kamionkowski:2014zda,Emami:2016mrt}. Within this context, the model considered here describes cosmic acceleration through the dynamics of a canonical scalar field evolving under a generalised axion-like potential, whose functional form extends the standard periodic axion potential. Constructed as a phenomenological framework, this potential supports an extended tracking regime that reduces the sensitivity to the initial scalar field conditions, followed by a late-time evolution approaching an effective cosmological constant.

%\Mariam{In this paper, we focus on a class of quintessence models inspired by axion-like dark energy scenarios, motivated by the framework introduced in \cite{Boiza:2024azh, Boiza:2024fmr}. In this approach, late-time cosmic acceleration is driven by the dynamics of a canonical scalar field evolving under a generalised axion-like potential, whose functional form extends the standard periodic axion potential in order to realise viable dark energy behaviour at late-time. The model is constructed at a phenomenological level and is designed to exhibit tracking properties and an effective cosmological constant behaviour in the recent Universe. Please Carlos, complete this paragraph  by adding a few sentence about axion-dark energy and a sentence on wave dark matter citing Tom's work.}

In this work, we extend the observational analysis of our previous study \cite{Chiang:2025qxg,Bouhmadi-Lopez:2025wxo} by performing a background-level parameter inference of the model using current cosmological data. The constraints are obtained through a Markov Chain Monte Carlo analysis based on the cosmic microwave background (CMB) shift parameters, recent baryon acoustic oscillation (BAO) measurements from DESI Data Release~2 \cite{DESI:2025zgx}, and luminosity distance data from the Pantheon+ type Ia supernovae (SNe) samples \cite{Brout:2022vxf} combined with the SH0ES determination of the Hubble constant \cite{Riess:2021jrx} or the cosmic chronometer (CC) based on galaxy cluster star burst history \cite{Moresco:2024wmr} and the luminosity data from the standardised gamma ray burst (GRB) \cite{Liu:2014vda}. This approach allows us to assess the consistency of the model with the observed expansion history of the Universe over a wide redshift range. These datasets naturally fall into two categories according to the Hubble parameter $H_0$: the higher $H_0$ branch with $H_0 \sim 73{\rm ~km/s/Mpc}$ representative in our work by the SH0ES project, and the lower $H_0$ ones with $H_0 \sim 67 {\rm ~km/s/Mpc}$, as suggested by CC and the combined constraint of CMB and BAO \footnote{Standardised candles such as type Ia SNe, GRB, and the acoustic scale derived from either CMB or BAO alone cannot pinpoint $H_0$.}. The aim of this work is to check if the axion-like dark energy model may alter this picture, by increasing the deceleration parameter during the tracking regime.

This paper is organised as follows. In Section~\ref{sec:gen_axion_summary}, we present the theoretical framework of the model and introduce the scalar field dynamics under consideration. Section~\ref{observations} describes the observational data sets employed in our analysis and outlines the methodology used to constrain the model parameters. In Section~\ref{fits}, we confront the theoretical predictions with current cosmological observations and discuss the resulting parameter constraints. Finally, in Section~\ref{conclusions}, we summarise our main findings and present our conclusions.

\section{Cosmology of generalised axion-like quintessence}
\label{sec:gen_axion_summary}

In this section, we briefly review the quintessence model introduced in~\cite{Boiza:2024azh}, which employs a generalised axion-like scalar field potential to account for the late-time accelerated expansion of the Universe. The model is formulated within a spatially flat Friedmann--Lemaître--Robertson--Walker (FLRW) framework and includes the standard radiation and pressureless matter components, the latter
comprising both cold dark matter and baryonic matter.

The scalar field dynamics are governed by the Klein--Gordon equation,
\begin{equation}
\ddot{\phi} + 3H\dot{\phi} + \frac{\mathrm{d}V(\phi)}{\mathrm{d}\phi} = 0 ,
\end{equation}
where $H \equiv \dot{a}/a$ denotes the Hubble expansion rate,
an overdot represents differentiation with respect to cosmic time,
and $V(\phi)$ is the scalar field potential.

The defining feature of the model is the axion-inspired potential
\begin{equation}
\label{eq:gen_axion_potential}
V(\phi) = \Lambda^4
\left[1 - \cos\!\left(\frac{\phi}{f}\right)\right]^{-n},
\qquad n>0,
\end{equation}
with $\Lambda$ and $f$ positive constants setting the energy scale and periodicity of the field. Treated as an effective phenomenological ansatz, this potential allows for an extended tracking phase in which the scalar field energy density follows that of the dominant background component, before driving the late-time accelerated expansion.

The background evolution is determined by the Friedmann equation,
\begin{equation}
H^2 = \frac{8\pi G}{3}\left(\rho_{\rm r} + \rho_{\rm m} + \rho_\phi \right),
\end{equation}
where $\rho_{\rm r}$ denotes the radiation energy density,
$\rho_{\rm m}=\rho_{\rm cdm}+\rho_{\rm b}$ is the total matter energy
density, with $\rho_{\rm cdm}$ and $\rho_{\rm b}$ corresponding to
cold dark matter and baryons, respectively, and $\rho_\phi$ is the
scalar field energy density, given by
\begin{equation}
\rho_\phi = \frac{1}{2}\dot{\phi}^2 + V(\phi).
\end{equation}
The corresponding scalar field pressure is given by
\begin{equation}
p_\phi = \frac{1}{2}\dot{\phi}^2 - V(\phi),
\end{equation}
and its equation of state parameter is defined as
\begin{equation}
w_\phi \equiv \frac{p_\phi}{\rho_\phi}
=
\frac{\frac{1}{2}\dot{\phi}^2 - V(\phi)}
{\frac{1}{2}\dot{\phi}^2 + V(\phi)}.
\end{equation}

The acceleration of the scale factor is given by
\begin{equation}
\frac{\ddot{a}}{a}
=
-\frac{4\pi G}{3}
\left(\rho_{\rm m}+2\rho_{\rm r}+\rho_\phi+3p_\phi\right).
\end{equation}
It is also useful to introduce the total effective equation
of state of the Universe, defined as
\begin{equation}
w_{\rm tot}
\equiv
\frac{p_{\rm tot}}{\rho_{\rm tot}}
=
\frac{\frac{1}{3}\rho_{\rm r}+p_\phi}
{\rho_{\rm r}+\rho_{\rm m}+\rho_\phi}
=
\frac{1}{3}\Omega_{\rm r}+w_\phi\Omega_\phi.
\label{eq:wtot}
\end{equation}
The acceleration equation can then be rewritten as
\begin{equation}
\frac{\ddot{a}}{a}
=
-\frac{H^2}{2}(1+3w_{\rm tot}),
\end{equation}
showing that accelerated expansion occurs when
$w_{\rm tot}<-1/3$.

Radiation and matter are minimally coupled and independently conserved,
\begin{align}
\dot{\rho}_{\rm r} + 4H\rho_{\rm r} &= 0 ,\\
\dot{\rho}_{\rm m} + 3H\rho_{\rm m} &= 0 .
\end{align}
Together, these equations fully specify the background cosmological dynamics of the model.

It is useful to introduce the usual density parameters
\begin{equation}
\Omega_A \equiv \frac{8\pi G}{3H^2}\rho_A,
\qquad A\in\{ {\rm r,m},\phi\},
\end{equation}
which satisfy the closure relation $\Omega_{\rm r}+\Omega_{\rm m}+\Omega_\phi=1$
in a spatially flat universe.

\begin{figure}[t]
  \centering
  \includegraphics[width=0.48\textwidth]{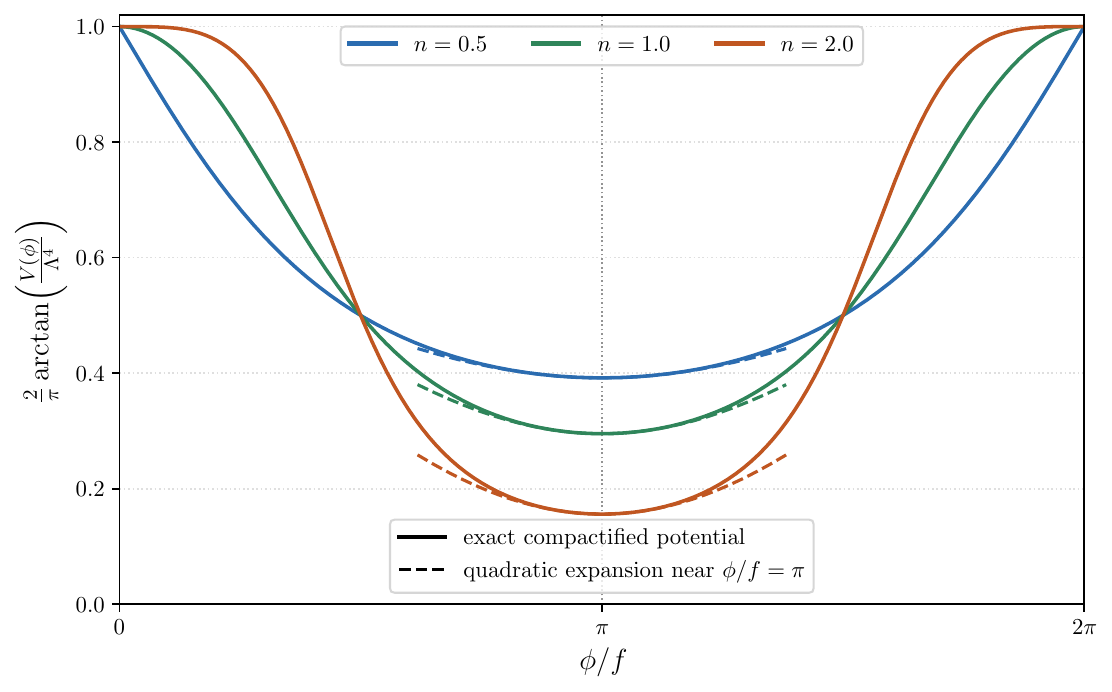}
  \caption{\justifying{Compactified generalised axion-like potential
  $V(\phi)/\Lambda^4=[1-\cos(\phi/f)]^{-n}$ as a function of $\phi/f$ over one period,
  for representative values of $n$. The vertical axis uses the compactified mapping
  $\frac{2}{\pi}\arctan\!\big(V/\Lambda^4\big)$, which renders the
  divergences at $\phi/f=0$ and $2\pi$ finite while preserving the minimum at
  $\phi/f=\pi$. Dashed curves show the quadratic expansion around the minimum,
  $V/\Lambda^4 \simeq 2^{-n} + \frac{n}{4}2^{-n}(\phi/f-\pi)^2$.}}
  \label{fig:axion_potential_norm}
\end{figure}

\subsection{Potential properties: minimum, vacuum scale, and effective mass}
\label{subsec:genaxion_min_mass}

The potential \eqref{eq:gen_axion_potential} reflects the periodic structure of
the cosine function. Owing to its inverse power-law form, it diverges at
$\phi/f = 2\pi k$, with $k\in\mathbb{Z}$, defining singular boundaries
that separate different branches of the potential. Within each branch,
the potential possesses a finite minimum at
\begin{equation}
\phi_{\rm min} = (2k+1)\pi f,
\qquad k\in\mathbb{Z},
\end{equation}
corresponding to $V_{,\phi}(\phi_{\rm min})=0$.
In this work, we restrict our analysis to the first branch,
$0<\phi/f<2\pi$, whose minimum is located at $\phi_{\rm min}=\pi f$.
This minimum defines the physically relevant late-time attractor
for the scalar field trajectories considered here.

The value of the potential at the minimum is
\begin{equation}
\label{eq:Vmin}
V_{\rm min} \equiv V(\phi_{\rm min}) = 2^{-n}\Lambda^4,
\end{equation}
which acts as an effective vacuum energy once the scalar field settles near
$\phi_{\rm min}$.

Expanding around the minimum and defining
$\Delta_\phi \equiv \phi - \phi_{\rm min}$,
the potential admits the quadratic approximation
\begin{equation}
V(\phi) \simeq
V_{\rm min}
+ \frac{1}{2} m_{\rm eff}^2 \Delta_\phi^2
+ \mathcal{O}(\Delta_\phi^3),
\end{equation}
which defines the effective mass (curvature) scale
\begin{equation}
\label{eq:meff}
m_{\rm eff}^2
\equiv
V_{,\phi\phi}\big|_{\phi=\phi_{\rm min}}
=
\frac{n}{2 f^2}\,V_{\rm min}
=
\frac{n}{2 f^2}\,2^{-n}\Lambda^4.
\end{equation}
This curvature scale determines the late-time dynamics of the scalar field
and, in particular, whether the approach to the minimum is overdamped
or oscillatory.

The global structure of the potential and the location of its finite minimum
are illustrated in Fig.~\ref{fig:axion_potential_norm}.

\subsection{Tracking behaviour and late-time approach to de~Sitter}
\label{subsec:genaxion_tracking}

A central feature of the model is the existence of an early-time tracking regime.
In the region of field space relevant during radiation and matter domination, the
potential behaves effectively as an inverse power-law. Indeed, for $|\phi|/f\ll1$,
\begin{equation}
1-\cos\!\left(\frac{\phi}{f}\right)\simeq \frac{1}{2}\left(\frac{\phi}{f}\right)^2,
\qquad\Rightarrow\qquad
V(\phi)\propto \phi^{-2n},
\end{equation}
so that the dynamics resemble those of inverse power-law quintessence.
As a result, $\rho_\phi$ can remain subdominant while evolving along an attractor-like
trajectory that is only weakly sensitive to initial conditions.

\begin{figure*}[t]
\centering

% --- Top row: equation of state ---
\begin{subfigure}[t]{0.80\textwidth}
    \centering
    \includegraphics[width=\linewidth]{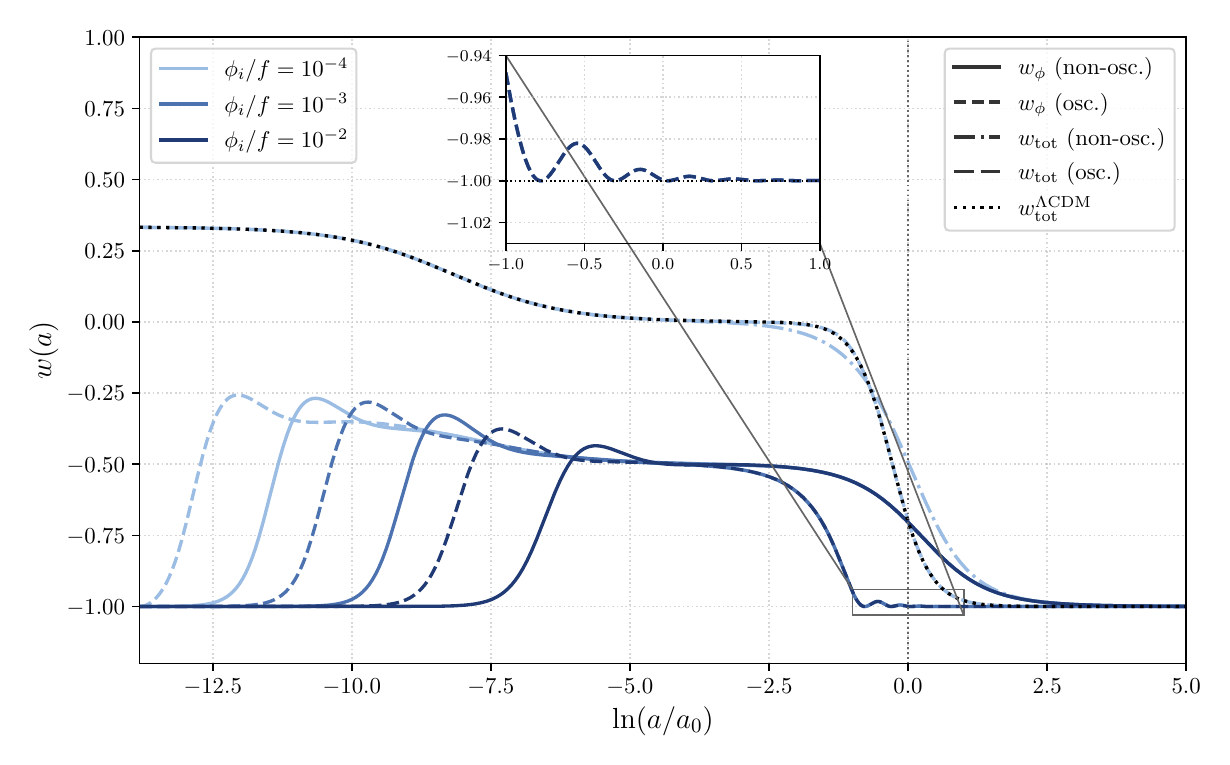}
    \caption{\justifying{Evolution of the scalar field equation of state
    $w_\phi$ and of the total effective equation of state $w_{\rm tot}$ for
    several initial displacements $\phi_{\rm ini}/f$, with
    $\dot{\phi}_{\rm ini}=0$. The values of $n$ and $f$ are kept fixed
    separately for the non-oscillatory and oscillatory branches, with
    $n=1$, $f=1\,M_{\rm Pl}$ and $f=0.1\,M_{\rm Pl}$, respectively.
    The different initial conditions mainly affect the time at which
    the field joins the tracker, while the trajectories subsequently
    converge toward the common matter-era tracking behaviour and
    the same late-time de~Sitter regime. The inset highlights the damped
    oscillations of $w_\phi$ around $-1$ in the oscillatory branch.}}
    \label{fig:axion_eos_wphi_wtot}
\end{subfigure}

\vspace{0.35cm}

% --- Bottom row: two panels ---
\begin{subfigure}[t]{0.49\textwidth}
    \centering
    \includegraphics[width=\linewidth]{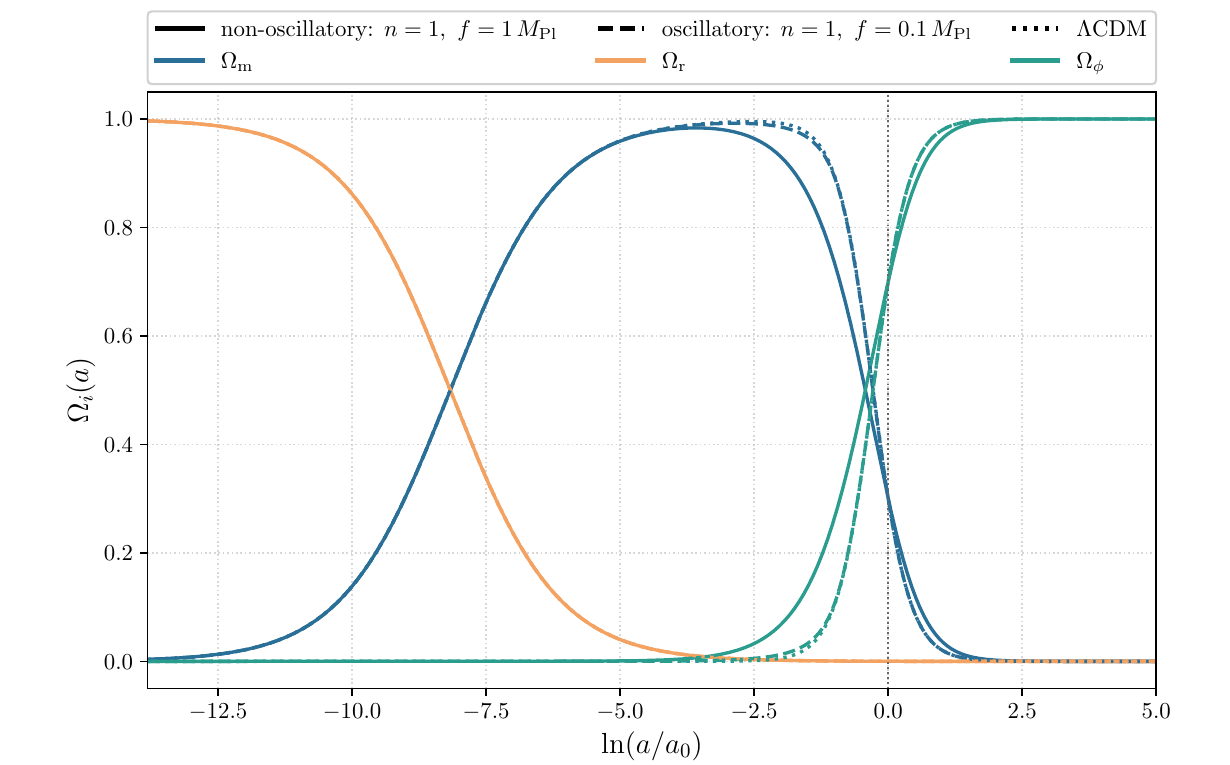}
    \caption{\justifying{Evolution of the density parameters
    $\Omega_{\rm r}$, $\Omega_{\rm m}$ and $\Omega_\phi$ for the generalised
    axion-like quintessence model with $n=1$. We compare a non-oscillatory
    trajectory with $f=1\,M_{\rm Pl}$ and an oscillatory trajectory with
    $f=0.1\,M_{\rm Pl}$, together with the corresponding $\Lambda$CDM
    reference. In both scalar field trajectories the initial conditions
    are fixed to $\phi_{\rm ini}/f=10^{-4}$ and
    $\dot{\phi}_{\rm ini}=0$.}}
    \label{fig:axion_omegas_evolution}
\end{subfigure}
\hfill
\begin{subfigure}[t]{0.49\textwidth}
    \centering
    \includegraphics[width=\linewidth]{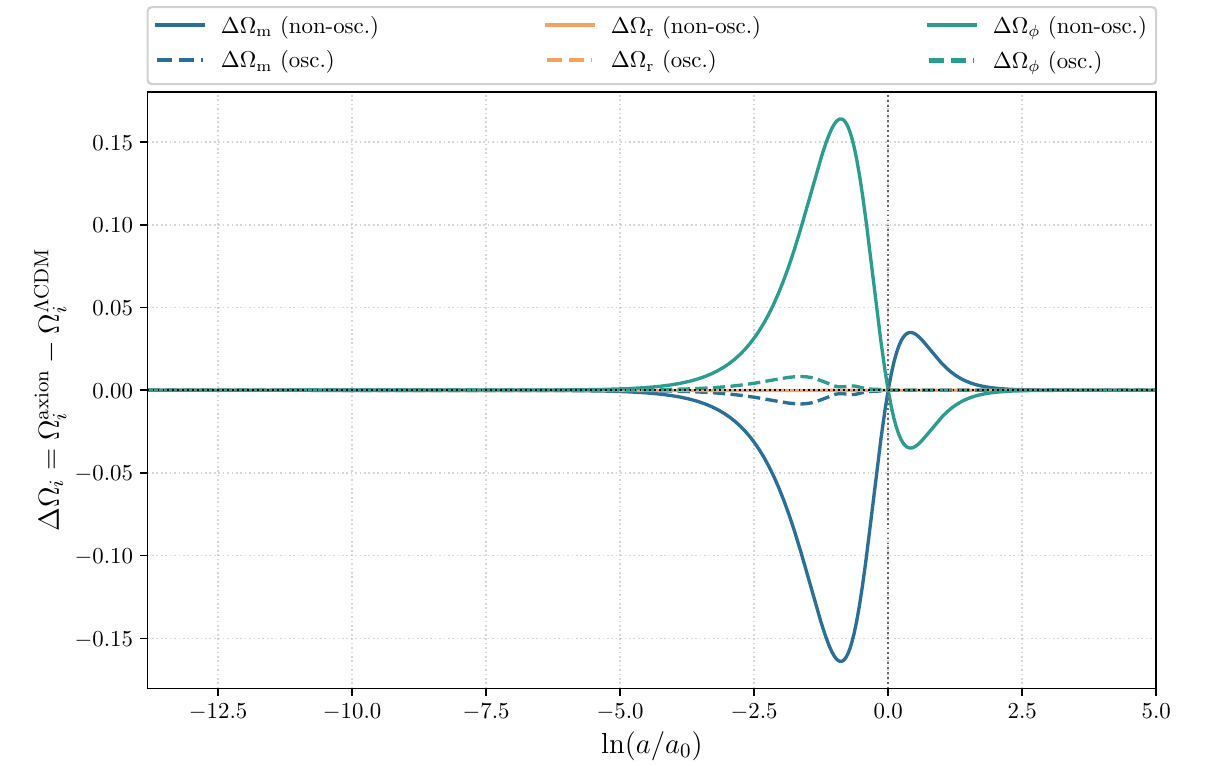}
    \caption{\justifying{Residuals of the density parameters with respect
    to $\Lambda$CDM,
    $\Delta\Omega_i\equiv\Omega_i^{\rm axion}-\Omega_i^{\Lambda{\rm CDM}}$,
    for $i=\{{\rm r},{\rm m},\phi\}$. The same parameter choices and initial
    conditions as in panel (b) are used, showing that the non-oscillatory
    branch produces larger late-time deviations from $\Lambda$CDM than the
    oscillatory branch.}}
    \label{fig:axion_omegas_residuals}
\end{subfigure}

\caption{\justifying{Representative background evolution of the generalised
axion-like quintessence model for $n=1$. Panel (a) tests the dependence on
the initial scalar field displacement by varying $\phi_{\rm ini}/f$ while
keeping the values of $n$ and $f$ fixed within each branch. The different
trajectories exhibit the expected tracking behaviour, approaching
$w_\phi\simeq -1/2$ during matter domination for $n=1$, and later evolving
toward the same asymptotic de~Sitter solution with $w_\phi\to -1$.
Panels (b) and (c) compare two late-time regimes: a non-oscillatory case
with $f=1\,M_{\rm Pl}$ and an oscillatory case with $f=0.1\,M_{\rm Pl}$,
both evolved from $\phi_{\rm ini}/f=10^{-4}$ and
$\dot{\phi}_{\rm ini}=0$. Panel (b) shows the radiation, matter and
scalar field density parameters, while panel (c) shows their residuals
with respect to the $\Lambda$CDM reference. The comparison shows that
the value of $f$ controls the late-time approach to the minimum of the
potential, distinguishing oscillatory from overdamped evolution, whereas
the tracker reduces the sensitivity to the initial value of the scalar field.}}
\label{fig:axion_background_summary}
\end{figure*}

During matter domination, the tracking solution is well approximated by
\cite{Boiza:2024azh}
\begin{equation}
\label{eq:wphi_tracker}
w_\phi \simeq -\frac{1}{1+n},
\qquad (\text{matter domination}),
\end{equation}
so that the scalar field energy density redshifts more slowly than matter and its
density fraction grows with time. For the representative case $n=1$ considered below,
this yields $w_\phi \simeq -1/2$, independently of the value of $f$. This reflects the
fact that the tracking dynamics are governed by the effective inverse power-law behaviour
of the potential at small field values, and are therefore insensitive to the curvature
scale near the minimum. As a result, both the oscillatory and non-oscillatory solutions exhibit the same
tracking behaviour prior to the onset of scalar field domination. This behaviour
is illustrated in Fig.~\ref{fig:axion_eos_wphi_wtot}, where varying
$\phi_{\rm ini}/f$ shows that the tracker is reached from a range of initial
scalar field displacements. This gradual growth
naturally triggers a transition to a late-time phase in which the scalar field becomes
dynamically important and drives accelerated expansion as it settles near the minimum.

\subsection{Oscillatory versus non-oscillatory late-time dynamics}
\label{subsec:genaxion_osc_criterion}

Once the field reaches the vicinity of the minimum, the displacement
$\Delta_\phi \equiv \phi - \phi_{\rm min}$
obeys, to leading order,
\begin{equation}
\ddot{\Delta}_\phi + 3H\dot{\Delta}_\phi + m_{\rm eff}^2\,\Delta_\phi \simeq 0,
\end{equation}
corresponding to a damped harmonic oscillator with Hubble friction.

Assuming that $H$ varies slowly over one oscillation period,
the characteristic exponents are
\begin{equation}
\lambda_\pm
=
-\frac{3H}{2}
\pm
\sqrt{\left(\frac{3H}{2}\right)^2 - m_{\rm eff}^2}.
\end{equation}
The motion becomes underdamped and oscillatory when
\begin{equation}
m_{\rm eff} > \frac{3}{2}H,
\end{equation}
whereas for $m_{\rm eff} < \frac{3}{2}H$ the evolution is overdamped and the
field approaches the minimum monotonically.

Since $H$ decreases with time, a transition from overdamped
to underdamped behaviour may occur dynamically as the Universe expands.
A practical late-time diagnostic is therefore the comparison with the
present expansion rate,
\begin{equation}
\label{eq:osc_diag}
\begin{aligned}
m_{\rm eff} \gtrsim H_0 
&\;\Rightarrow\; \text{oscillations can occur by today}, \\
m_{\rm eff} \lesssim H_0 
&\;\Rightarrow\; \text{overdamped approach}.
\end{aligned}
\end{equation}

Using the expression
\begin{equation}
m_{\rm eff}^2 = \frac{n}{2 f^2} V_{\rm min},
\end{equation}
and assuming that the field reaches the minimum during the onset
of scalar field domination, so that
\begin{equation}
3H^2 \simeq 8\pi G\,V_{\rm min},
\end{equation}
one obtains the approximate relation
\begin{equation}
m_{\rm eff}^2
\simeq
\frac{3n}{16\pi G\,f^2}\,H^2.
\end{equation}
The condition $m_{\rm eff} = 3H/2$ then yields a characteristic
critical scale
\begin{equation}
\label{eq:fcrit}
f_{\rm crit}
\equiv
\sqrt{\frac{n}{12\pi G}}
=
\sqrt{\frac{2n}{3}}\, M_{\rm Pl},
\end{equation}
where $M_{\rm Pl} \equiv (8\pi G)^{-1/2}$.
Therefore, oscillatory behaviour is expected when
$f \lesssim f_{\rm crit}$,
whereas for $f \gtrsim f_{\rm crit}$
the evolution remains overdamped.

Regardless of whether the late-time evolution is overdamped or oscillatory,
both branches converge toward the same asymptotic de~Sitter solution
characterised by
\begin{equation}
H^2 \rightarrow \frac{8\pi G}{3}\,V_{\rm min},
\qquad
w_\phi \rightarrow -1.
\end{equation}
The distinction between the two regimes therefore concerns only the
dynamical path by which the scalar field approaches the minimum,
while the final background state is identical.

The two late-time regimes are illustrated in Fig.~\ref{fig:axion_background_summary}
for the representative case $n=1$. In Figs.~\ref{fig:axion_omegas_evolution} and \ref{fig:axion_omegas_residuals} we compare an oscillatory
trajectory, with $f=0.1\,M_{\rm Pl}$, and a non-oscillatory trajectory, with
$f=1\,M_{\rm Pl}$, fixing in both cases the same initial conditions,
$\phi_{\rm ini}/f=10^{-4}$ and $\dot{\phi}_{\rm ini}=0$. In the oscillatory case,
the field reaches the vicinity of the minimum earlier and the subsequent
oscillations are efficiently damped by the Hubble expansion. As a result, the
scalar field rapidly approaches the de~Sitter attractor and the background
evolution remains very close to that of $\Lambda$CDM, with only mild late-time
departures in the density parameters. By contrast, in the non-oscillatory case,
the approach to the minimum is overdamped and more gradual, leading to more
visible deviations from $\Lambda$CDM in the recent expansion history.

Fig.~\ref{fig:axion_eos_wphi_wtot} complements this comparison by varying the initial displacement
$\phi_{\rm ini}/f$, while keeping the values of $n$ and $f$ fixed separately in
the oscillatory and non-oscillatory cases. This makes the tracking behaviour
explicit: different initial conditions may delay or advance the moment at which
the field joins the tracker, but once the tracking regime is reached the
trajectories evolve toward the same matter-era behaviour and subsequently toward
the same late-time de~Sitter limit. Thus, the late-time evolution is controlled
primarily by the structure of the potential and by the value of $f$, rather than
by a fine tuning of the initial scalar field displacement.

%====================================================

\section{Observational data}
\label{observations}

In this section we describe the cosmological datasets used to constrain the
generalised axion-like quintessence model. Since the analysis is restricted to
the homogeneous background evolution, we employ probes that primarily measure
cosmological distances or the expansion rate over a broad range of redshifts.
The datasets considered here are CMB distance priors, baryon acoustic
oscillation measurements from DESI DR2, Type Ia supernovae from Pantheon+,
local measurements of the Hubble constant from SH0ES, cosmic chronometer
measurements of $H(z)$, and gamma-ray burst distance data.

These observations provide complementary information on different epochs of the
cosmic expansion history. The CMB distance priors constrain the geometry of the
Universe close to photon decoupling. BAO measurements use the sound horizon at
the baryon drag epoch as a standard ruler and therefore provide absolute
distance information at low and intermediate redshifts. Type Ia supernovae trace
the relative luminosity-distance relation at late-time, while the SH0ES
calibration fixes the corresponding absolute scale. Cosmic chronometers give
direct estimates of the expansion rate, and gamma-ray bursts extend the Hubble
diagram to redshifts beyond those typically reached by supernovae.

For the purposes of the present analysis, we organise the data into three
classes. The first class is a high-redshift set formed by CMB distance priors
and BAO measurements. The second class is a high-$H_0$ late-time distance-ladder
set, consisting of the Pantheon+ supernova compilation calibrated with SH0ES
Cepheid measurements, which we denote by PPS. The third class is a lower-$H_0$
late-time set formed by cosmic chronometers and gamma-ray bursts, together with
a binned Pantheon+ supernova sample without the SH0ES calibration, which we
denote by Pan.

We analyse the following dataset combinations:
\begin{enumerate}
    \item BAO;
    \item CMB + BAO;
    \item PPS;
    \item CC + GRB + Pan;
    \item CMB + BAO + PPS;
    \item CMB + BAO + CC + GRB + Pan.
\end{enumerate}
We do not consider the CMB distance priors as an independent dataset in the
parameter inference. The compressed CMB likelihood contains only three effective
observables, which is fewer than the number of free parameters varied in the
model.

\subsection{CMB distance priors}
\label{subsec:cmb_data}

The Cosmic Microwave Background contains precise information about the geometry
and physical conditions of the Universe at the time of photon decoupling. For
models in which the main deviations from $\Lambda$CDM occur at late-time, much
of the relevant background information can be encoded in a small set of
compressed variables known as distance priors or shift parameters.

We use the CMB distance priors derived in~\cite{Zhai:2018vmm}. The data
vector is given by
\begin{equation}
X_{\rm CMB}
=
\left(
R,\ell_a,\omega_{\rm b0}
\right),
\end{equation}
where $R$ is the shift parameter, $\ell_a$ is the acoustic angular scale, and
\begin{equation}
\omega_{\rm b0}
=
\Omega_{\rm b0}h^2
\end{equation}
is the present physical baryon density. Here $h$ is defined through
\begin{equation}
H_0
=
100h\ {\rm km\,s^{-1}\,Mpc^{-1}}.
\end{equation}

The shift parameter is sensitive to the distance to the surface of last
scattering weighted by the present matter density, whereas the acoustic scale
measures the angular size of the sound horizon at photon decoupling. The baryon
density controls the baryonic contribution to the pre-recombination plasma and
enters the calculation of both the sound horizon and the decoupling redshift.
In the likelihood analysis we use the covariance matrix associated with these
three quantities, thereby retaining the correlations among them.

\subsection{DESI DR2 BAO measurements}
\label{subsec:bao_data}

Baryon acoustic oscillations originate from acoustic waves in the tightly
coupled photon-baryon fluid before recombination. After photons decouple, the
same physical scale remains imprinted in the late-time distribution of matter.
This scale is set by the comoving sound horizon at the baryon drag epoch,
$r_{\rm d}$, and can be used as a standard ruler for cosmological distance
measurements.

We employ the full BAO dataset from the second data release of DESI
\cite{DESI:2025zpo,DESI:2025zgx}. This compilation contains 13 measurements
covering redshifts up to $z=2.33$. Depending on the tracer and redshift bin, the
observational constraints are reported in terms of the dimensionless ratios
\begin{equation}
\frac{D_M(z)}{r_{\rm d}},
\qquad
\frac{D_H(z)}{r_{\rm d}},
\qquad
\frac{D_V(z)}{r_{\rm d}}.
\end{equation}
Here $D_M(z)$ is the transverse comoving distance, $D_H(z)$ is the Hubble
distance, and $D_V(z)$ is the volume-averaged distance. These quantities probe,
respectively, the angular BAO scale, the radial BAO scale, and the isotropic BAO
scale. The DESI DR2 measurements therefore provide a precise reconstruction of
the expansion history over a wide redshift interval.

\subsection{Type Ia supernovae}
\label{subsec:sne_data}

Type Ia supernovae are standardisable candles and provide direct information on
the luminosity-distance relation. Their absolute luminosity is not known a
priori, but empirical corrections based on light-curve properties allow them to
trace relative distances with high precision.

We use the Pantheon+ compilation \cite{Brout:2022vxf}, which contains 1701
supernova light curves in the approximate redshift range
\begin{equation}
0.001 < z < 2.26 .
\end{equation}
In this work the Pantheon+ sample is used in two different ways. First, we
consider the Pantheon+ and SH0ES calibrated compilation, which combines the
supernova data with Cepheid-based distance-ladder information. We denote this
dataset by PPS. Since the SH0ES calibration fixes the absolute magnitude of
Type Ia supernovae, this dataset provides an absolute late-time distance probe
and is associated with a relatively high value of the Hubble constant.

Second, we also consider a Pantheon+ supernova sample without the SH0ES
calibration, denoted by Pan. In the absence of an external calibration, the
supernovae constrain the shape of the luminosity-distance relation but not its
absolute normalisation. For this case, we use the binned version of the
Pantheon+ sample, which consists of 40 data points and provides a compressed
description of the supernova Hubble diagram. This binned, uncalibrated
supernova information is combined with other probes, such as cosmic
chronometers and gamma-ray bursts, when constructing the lower-$H_0$ late-time
dataset.

\subsection{Cosmic chronometers}
\label{subsec:cc_data}

Cosmic chronometers provide direct measurements of the Hubble expansion rate
$H(z)$. The method is based on the differential ageing of passively evolving
galaxies. By estimating the variation of cosmic time with redshift, one obtains
\begin{equation}
H(z)
=
-\frac{1}{1+z}
\frac{dz}{dt}.
\end{equation}
This approach gives expansion rate measurements that do not rely on the
integrated distance-redshift relation.

We adopt the cosmic chronometer compilation of~\cite{Moresco:2024wmr}.
The sample contains
\begin{equation}
N_{\rm CC}=33
\end{equation}
measurements covering approximately
\begin{equation}
0.07 \lesssim z \lesssim 2.0 .
\end{equation}
These data are used as part of the CC + GRB + Pan dataset combination.

\subsection{Gamma-ray bursts}
\label{subsec:grb_data}

Gamma-ray bursts can be used as high-redshift distance indicators. Although they are less standardised than Type Ia supernovae, they are valuable
because they extend the Hubble diagram to significantly higher redshifts.

We use the Mayflower gamma-ray burst sample of~\cite{Liu:2014vda}. This
compilation contains
\begin{equation}
N_{\rm GRB}=79
\end{equation}
objects in the redshift interval
\begin{equation}
1.44 < z < 8.1 .
\end{equation}
The GRB data therefore complement the supernova and cosmic chronometer
measurements by adding distance information at redshifts where late-time probes
are otherwise sparse.

%====================================================

\section{Statistical methodology}
\label{sec:methodology}

We constrain the parameters of the generalised axion-like quintessence model
within a Bayesian framework. Given a model $M$, a set of parameters
$\boldsymbol{\theta}_M$, and a dataset $\mathcal{D}$, Bayes' theorem gives
\begin{equation}
P(\boldsymbol{\theta}_M|\mathcal{D},M)
=
\frac{
P(\mathcal{D}|\boldsymbol{\theta}_M,M)
P(\boldsymbol{\theta}_M|M)
}{
P(\mathcal{D}|M)
}.
\label{eq:bayes}
\end{equation}
Here $P(\boldsymbol{\theta}_M|\mathcal{D},M)$ is the posterior probability
distribution, $P(\mathcal{D}|\boldsymbol{\theta}_M,M)$ is the likelihood, and
$P(\boldsymbol{\theta}_M|M)$ is the prior probability distribution. The Bayesian
evidence \footnote{In the numerical analysis, the likelihood entering the posterior
also includes the auxiliary Gaussian constraints described in
Section~\ref{sec:prior}, namely the BBN determination of the
physical baryon density and, for the generalised axion-like model,
the present-day Friedmann consistency condition. These constraints
are retained throughout parameter inference, while their direct
contributions are excluded from the observational likelihood
employed in the model comparison and dataset concordance
diagnostics. The corresponding conditional Bayesian evidences
are defined in Section~\ref{sec:model-comparison}.},
\begin{equation}
P(\mathcal{D}|M)
=
\int
P(\mathcal{D}|\boldsymbol{\theta}_M,M)
P(\boldsymbol{\theta}_M|M)
d\boldsymbol{\theta}_M,
\end{equation}
acts as the normalisation of the posterior.

The posterior distributions are sampled using the Markov Chain Monte Carlo
sampler implemented in \texttt{Cobaya}
\cite{Torrado:2020dgo,Lewis:2002ah,Lewis:2013hha,Neal:2005uqf}. The resulting
chains are analysed and visualised with \texttt{GetDist}
\cite{Lewis:2019xzd}. For each cosmological model and dataset combination, we
run between two and four independent MCMC chains in order to assess convergence
and the numerical stability of the posterior constraints. Convergence is
monitored using the Gelman--Rubin criterion \cite{Lewis:2013hha}, which we evaluate both within each
individual chain and across independent chains \footnote{The quantity $R-1$ is defined as the largest eigenvalue of
$SCS^{\rm T}$, where $C$ is the covariance matrix of the mean parameter values
and $S$ is a matrix satisfying $SMS^{\rm T}=I$, with $M$ the mean covariance
matrix of the sampled parameters. For the inter-chain criterion, after removing
the earliest $30\%$ of samples as burn-in, we compute the mean vector and
covariance matrix of the sampled parameters for each chain. The matrix $C$ is
then obtained from the covariance of the chain mean vectors, while $M$ is
obtained by averaging the within-chain covariance matrices over all chains with
equal weight. For the intra-chain criterion, the first $20\%$ of each chain is
removed before burn-in removal, the remaining samples are divided into four
segments, and the same construction is applied by treating the four segments as
independent chains.}, requiring
\begin{equation}
R-1<0.1
\end{equation}
for intra-chain convergence and
\begin{equation}
R-1<0.01
\end{equation}
for inter-chain convergence. After these convergence criteria are satisfied, we discard the earliest $30\%$
of samples in each chain as burn-in. The remaining samples from all chains are
then combined and used to compute the marginalised constraints and derived
quantities.

\subsection{Likelihood functions}
\label{sec:likelihoods}

For all datasets considered in this work, we assume Gaussian likelihoods. The
likelihood of a dataset can therefore be written in terms of its chi-squared
statistic as
\begin{equation}
P(\mathcal{D}|\boldsymbol{\theta}_M,M)
\propto
\exp\left(
-\frac{\chi^2}{2}
\right).
\end{equation}
When several statistically independent datasets are combined, the total
likelihood is the product of the individual likelihoods. Equivalently, the total
chi-squared is obtained by summing the relevant contributions. For the dataset combinations considered in this work, the total chi-squared is
constructed by summing the statistically independent contributions included in
each case. In particular, the late-time sector is taken either from the
Pantheon+ \& SH0ES calibrated sample or from the combination
CC + GRB + Pan, but the two are not used simultaneously. Thus, schematically,
\begin{equation}
\chi^2_{\rm obs}
=
\chi^2_{\rm CMB}
+
\chi^2_{\rm BAO}
+
\chi^2_{\rm late},
\label{eq:chi2_total}
\end{equation}
where the relevant terms are included only when present in the dataset
combination, and
\begin{equation}
\chi^2_{\rm late}
=
\begin{cases}
\chi^2_{\rm PPS},
&
\begin{array}{l}
\text{for PPS} \\
\text{combinations},
\end{array}
\\[2mm]
\chi^2_{\rm CC}+\chi^2_{\rm GRB}+\chi^2_{\rm Pan},
&
\begin{array}{l}
\text{for CC + GRB + Pan} \\
\text{combinations}.
\end{array}
\end{cases}
\label{eq:chi2_late}
\end{equation}

\subsubsection{CMB distance-prior likelihood}

The theoretical CMB distance-prior vector is
\begin{equation}
X_{\rm CMB}^{\rm th}
=
\left(
R,\ell_a,\omega_{\rm b0}
\right).
\end{equation}
The shift parameter is defined as
\begin{equation}
R(z_*)
=
\frac{
\sqrt{\Omega_{\rm m0}H_0^2}
}{
c
}
(1+z_*)D_A(z_*),
\label{eq:R_cmb}
\end{equation}
where $z_*$ is the redshift of photon decoupling and $D_A(z)$ is the angular
diameter distance.

The acoustic angular scale is
\begin{equation}
\ell_a(z_*)
=
\pi(1+z_*)
\frac{
D_A(z_*)
}{
r_s(z_*)
},
\label{eq:la_cmb}
\end{equation}
where $r_s(z_*)$ is the comoving sound horizon at photon decoupling.

For the spatially flat cosmologies studied here, the angular diameter distance
is
\begin{equation}
D_A(z)
=
\frac{c}{1+z}
\int_0^z
\frac{dz'}{H(z')}.
\label{eq:angular_diameter_distance}
\end{equation}
The comoving sound horizon is computed as
\begin{equation}
r_s(z)
=
\int_z^\infty
\frac{c\,dz'}{H(z')}
\frac{1}{
\sqrt{
3\left[
1+
R_{\rm b}
(1+z')^{-1}
\right]
}
},
\label{eq:sound_horizon}
\end{equation}
where \cite{Eisenstein:1997ik}
\begin{equation}
R_{\rm b}
=
31500\,\omega_{\rm b0}
\left(
\frac{
T_{\rm CMB}
}{
2.7\,{\rm K}
}
\right)^{-4}.
\end{equation}
We fix the CMB temperature to \cite{Fixsen:2009ug}
\begin{equation}
T_{\rm CMB}
=
2.7255\,{\rm K}.
\end{equation}
The photon decoupling redshift is evaluated using the fitting formula
\cite{Eisenstein:1997ik,Hu:1995en}
\begin{equation}
z_*
=
1048
\left(
1+
0.00124\,\omega_{\rm b0}^{-0.738}
\right)
\left(
1+
g_1\omega_{\rm m0}^{g_2}
\right),
\end{equation}
with
\begin{equation}
g_1
=
0.0783\,\omega_{\rm b0}^{-0.238}
\left(
1+
39.5\,\omega_{\rm b0}^{0.763}
\right)^{-1},
\end{equation}
and
\begin{equation}
g_2
=
0.560
\left(
1+
21.1\,\omega_{\rm b0}^{1.81}
\right)^{-1}.
\end{equation}
The physical matter density is
\begin{equation}
\omega_{\rm m0}
=
\Omega_{\rm m0}h^2.
\end{equation}

The residual vector is defined as
\begin{equation}
\Delta X_{\rm CMB}
=
X_{\rm CMB}^{\rm obs}
-
X_{\rm CMB}^{\rm th},
\end{equation}
and the corresponding chi-squared is
\begin{equation}
\chi^2_{\rm CMB}
=
\Delta X_{\rm CMB}^{\rm T}
C_{\rm CMB}^{-1}
\Delta X_{\rm CMB},
\label{eq:chi2_cmb}
\end{equation}
where $C_{\rm CMB}$ is the covariance matrix of the CMB distance priors.

\subsubsection{DESI DR2 BAO likelihood}

The DESI DR2 BAO likelihood is constructed from measurements of
\begin{equation}
\frac{D_M(z)}{r_{\rm d}},
\qquad
\frac{D_H(z)}{r_{\rm d}},
\qquad
\frac{D_V(z)}{r_{\rm d}}.
\end{equation}
The Hubble distance is
\begin{equation}
D_H(z)
=
\frac{c}{H(z)},
\end{equation}
and the transverse comoving distance is
\begin{equation}
D_M(z)
=
(1+z)D_A(z).
\end{equation}
The volume-averaged distance is defined by
\begin{equation}
D_V(z)
=
\left[
zD_M^2(z)D_H(z)
\right]^{1/3}.
\end{equation}

The sound horizon at the drag epoch is calculated using the fitting expression
\begin{equation}
r_{\rm d}
=
147.05\,{\rm Mpc}
\left(
\frac{
\omega_{\rm b0}
}{
0.02236
}
\right)^{-0.13}
\left(
\frac{
\omega_{\rm m0}
}{
0.1432
}
\right)^{-0.23}.
\label{eq:rd}
\end{equation}
The original fitting relation also includes an additional correction factor
depending on the effective number of relativistic species, $N_{\rm eff}$.
Throughout our analysis we fix this parameter to its standard value,
$N_{\rm eff}=3.04$, so that the corresponding factor is equal to unity and is
not written explicitly in~\eqref{eq:rd}.

The BAO residual vector is
\begin{equation}
\Delta X_{\rm BAO}
=
X_{\rm BAO}^{\rm obs}
-
X_{\rm BAO}^{\rm th}.
\end{equation}
The BAO chi-squared is then
\begin{equation}
\chi^2_{\rm BAO}
=
\Delta X_{\rm BAO}^{\rm T}
C_{\rm BAO}^{-1}
\Delta X_{\rm BAO},
\label{eq:chi2_bao}
\end{equation}
where $C_{\rm BAO}$ denotes the DESI DR2 covariance matrix.

\subsubsection{Pantheon+ and SH0ES likelihood}
\label{subsubsec:pps_likelihood}

We use Type Ia supernovae from the Pantheon+ compilation calibrated with the
SH0ES distance ladder. The data are effectively given in terms of the observed
apparent magnitude, heliocentric redshift, and CMB-frame redshift,
$(m_i,z_{{\rm hel},i},z_{{\rm CMB},i})$. The observational distance modulus is
defined as
\begin{equation}
\mu_{{\rm SN},i}^{\rm obs}
=
m_i-M,
\end{equation}
where $M$ is the standardised absolute magnitude of Type Ia supernovae.

The theoretical distance modulus is obtained from
the Hubble-free luminosity distance as
\begin{equation}
\mu_{\rm SN}^{\rm th}
(z_{\rm hel},z_{\rm CMB})
=
5\log_{10}
\left[
D^{\rm free}_L(z_{\rm hel},z_{\rm CMB})
\right]
+
\mu_0,
\end{equation}
where
\begin{equation}
\mu_0
\equiv
5\log_{10}
\left[
cH_0^{-1}\,{\rm Mpc}^{-1}
\right]
+
25
\end{equation}
collects the unit conversion from Mpc to the reference distance of $10\,{\rm pc}$.
The dimensionless Hubble-free luminosity distance is
\begin{equation}
D^{\rm free}_L(z_{\rm hel},z_{\rm CMB})
=
(1+z_{\rm hel})
\int_0^{z_{\rm CMB}}
\frac{H_0}{H(z')}
\,{\rm d}z' .
\end{equation}

The Pantheon+ \& SH0ES compilation includes a set of Cepheid-host supernovae
that calibrate the distance scale through the SH0ES distance ladder. For
Hubble-flow supernovae, the standard cosmological prediction above is used. For
Cepheid-host supernovae, the distance entering the likelihood is instead
replaced by the measured Cepheid-based host distance modulus. This calibration
is incorporated through the full Pantheon+ \& SH0ES covariance matrix, which
accounts for statistical and systematic uncertainties as well as correlations
among supernovae.

Defining the residual vector
\begin{equation}
\Delta\mu_{\rm SN}
=
\mu_{\rm SN}^{\rm obs}
-
\mu_{\rm SN}^{\rm th},
\end{equation}
the Pantheon+ \& SH0ES chi-squared is written as
\begin{equation}
\chi^2_{\rm PPS}
=
\Delta\mu_{\rm SN}^{\rm T}
C_{\rm PPS}^{-1}
\Delta\mu_{\rm SN},
\label{eq:chi2_pps}
\end{equation}
where $C_{\rm PPS}$ is the covariance matrix of the Pantheon+ \& SH0ES dataset.

\subsubsection{Uncalibrated Pantheon+ likelihood}
\label{subsubsec:pan_likelihood}

For the Pan dataset, we use the binned Pantheon+ sample without the SH0ES
Cepheid calibration. This compressed sample consists of 40 measurements of the
supernova distance modulus and provides a binned representation of the
Pantheon+ Hubble diagram. The observed distance modulus of each binned supernova point is defined, as
previously, as
\begin{equation}
\mu_{{\rm SN},i}^{\rm obs}
=
m_i-M,
\end{equation}
where $m_i$ is the observed apparent magnitude of the $i$th binned data point
and $M$ is the standardised absolute magnitude of Type Ia supernovae. For a given cosmological model, the theoretical
prediction is computed from the dimensionless Hubble-free luminosity distance
defined above,
\begin{equation}
\mu_{\rm SN}^{\rm th}
(z_{\rm hel},z_{\rm CMB})
=
5\log_{10}
\left[
D^{\rm free}_L(z_{\rm hel},z_{\rm CMB})
\right]
+
\mu_0,
\end{equation}
with the same definition of $\mu_0$ as in the Pantheon+ \& SH0ES case.

Defining the residual vector
\begin{equation}
\Delta\mu_{\rm SN}
=
\mu_{\rm SN}^{\rm obs}
-
\mu_{\rm SN}^{\rm th},
\end{equation}
the corresponding chi-squared statistic is
\begin{equation}
\chi^2_{\rm Pan}
=
\Delta\mu_{\rm SN}^{\rm T}
C_{\rm Pan}^{-1}
\Delta\mu_{\rm SN},
\end{equation}
where $C_{\rm Pan}$ is the covariance matrix of the binned Pantheon+ sample.

Since the uncalibrated binned Pantheon+ sample does not fix the absolute magnitude scale, the likelihood depends on the nuisance parameter $\widetilde{M}\equiv M+\mu_0$. Equivalently, the residual can be written as 
\begin{equation} 
\mu_{{\rm SN},i}^{\rm obs} - \mu_{{\rm SN},i}^{\rm th} = m_i - 5\log_{10} \left[ D_L^{\rm free}(z_{{\rm hel},i},z_{{\rm CMB},i}) \right] - \widetilde{M}. 
\end{equation} 
We marginalise analytically over $\widetilde{M}$. Following~\cite{SNLS:2011lii}, the resulting marginalised statistic is 
\begin{equation} 
\widetilde{\chi}^2_{\rm Pan} = A_{\rm Pan} + \ln\!\left(\frac{E_{\rm Pan}}{2\pi}\right) - \frac{B_{\rm Pan}^2}{E_{\rm Pan}}, \label{eq:chi2_pan_marg} 
\end{equation} 
where 
\begin{equation} 
\begin{aligned} A_{\rm Pan} &= \Delta\widetilde{\mu}_{\rm SN}^{\rm T} C_{\rm Pan}^{-1} \Delta\widetilde{\mu}_{\rm SN}, \\ B_{\rm Pan} &= \Delta\widetilde{\mu}_{\rm SN}^{\rm T} C_{\rm Pan}^{-1} 1, \\ E_{\rm Pan} &= 1^{\rm T} C_{\rm Pan}^{-1} 1, \\
\Delta\widetilde{\mu}_{\rm SN}
&=
\Delta\mu_{\rm SN}(\widetilde{M}=0).
\end{aligned}
\end{equation}
In the present analysis, the uncalibrated binned Pantheon+ sample is always
combined with cosmic chronometer measurements, which provide an independent
calibration of the expansion scale. This dataset combination is also always
used together with the GRB Hubble-diagram sample.

\subsubsection{Cosmic chronometer likelihood}

The cosmic chronometer data provide direct measurements of $H(z)$. For each
redshift $z_i$ in the compilation, the theoretical prediction is obtained by
evaluating the background expansion rate of the model at that redshift. Assuming
uncorrelated measurements, the chi-squared is
\begin{equation}
\chi^2_{\rm CC}
=
\sum_{i=1}^{N_{\rm CC}}
\frac{
\left[
H^{\rm obs}(z_i)
-
H^{\rm th}(z_i)
\right]^2
}{
\sigma_{H,i}^2
},
\label{eq:chi2_cc}
\end{equation}
where $\sigma_{H,i}$ is the uncertainty of the $i$th measurement.

\subsubsection{Gamma-ray burst likelihood}
\label{subsubsec:grb_likelihood}

The GRB dataset is provided in terms of the observed distance modulus
$\mu_{\rm GRB}^{\rm obs}$ at a given redshift, together with the corresponding
uncertainty. For a given cosmological model, we compute the theoretical
prediction $\mu_{\rm GRB}^{\rm th}(z)$ from the luminosity distance in a way
analogous to the one used for Type Ia supernovae, except that there is no need
to distinguish between heliocentric and CMB-frame redshifts. Thus, we use
\begin{equation}
\mu_{\rm GRB}^{\rm th}(z)
=
5\log_{10}
\left[
D_L^{\rm free}(z)
\right]
+
\mu_0,
\end{equation}
where $D_L^{\rm free}(z)$ denotes the dimensionless Hubble-free luminosity
distance.

Defining the residual vector
\begin{equation}
\Delta\mu_{\rm GRB}
=
\mu_{\rm GRB}^{\rm obs}
-
\mu_{\rm GRB}^{\rm th},
\end{equation}
the corresponding chi-squared statistic is
\begin{equation}
\chi^2_{\rm GRB}
=
\Delta\mu_{\rm GRB}^{\rm T}
C_{\rm GRB}^{-1}
\Delta\mu_{\rm GRB},
\end{equation}
where $C_{\rm GRB}$ is the covariance matrix of the GRB sample. In this work,
we assume this matrix to be diagonal,
$C_{\rm GRB}=\mathrm{diag}(\sigma_{\mu,i}^2)$, with $\sigma_{\mu,i}$ the
reported uncertainty of the $i$th data point.

Since $\mu_{\rm GRB}^{\rm th}$ depends on the nuisance parameter $\mu_0$, we
marginalise analytically over $\mu_0$. Following~\cite{SNLS:2011lii}, the
resulting marginalised statistic is
\begin{equation}
\widetilde{\chi}^2_{\rm GRB}
=
A_{\rm GRB}
+
\ln\!\left(\frac{E_{\rm GRB}}{2\pi}\right)
-
\frac{B_{\rm GRB}^2}{E_{\rm GRB}},
\label{eq:chi2_grb_marg}
\end{equation}
where
\begin{equation}
\begin{aligned}
A_{\rm GRB}
&=
\Delta\widetilde{\mu}_{\rm GRB}^{\rm T}
C_{\rm GRB}^{-1}
\Delta\widetilde{\mu}_{\rm GRB},
\\
B_{\rm GRB}
&=
\Delta\widetilde{\mu}_{\rm GRB}^{\rm T}
C_{\rm GRB}^{-1}
1,
\\
E_{\rm GRB}
&=
1^{\rm T}
C_{\rm GRB}^{-1}
1,
\\
\Delta\widetilde{\mu}_{\rm GRB}
&=
\Delta\mu_{\rm GRB}(\mu_0=0).
\end{aligned}
\end{equation}

\subsection{Parameter space and prior specification}
\label{sec:prior}

The parameter space explored in this work consists of the standard late-time cosmological parameters together with a set of quantities describing the scalar potential of the generalised axion-like dark energy model. The cosmological parameters are the Hubble constant $H_0$ and the present-day matter and baryon density parameters $\Omega_{\rm m0}$ and $\Omega_{\rm b0}$ \footnote{To compute the present radiation density parameter $\Omega_{\rm r0}$ we first estimate the redshift of matter--radiation equality using $z_{\rm eq}=2.5\times10^4\,\Omega_{\rm m0}h^2\left(\frac{T_{\rm CMB}}{2.7\,{\rm K}}\right)^{-4}$ \cite{Eisenstein:1997ik}, with $T_{\rm CMB}=2.7255\,{\rm K}$ \cite{Fixsen:2009ug}. The radiation density is then fixed through $\Omega_{\rm r0}=\Omega_{\rm m0}/(1+z_{\rm eq})$. Assuming a spatially flat Universe, the dark-energy density today follows from $\Omega_{\phi0}=1-\Omega_{\rm m0}-\Omega_{\rm r0}$.}. The additional parameters determine the shape of the scalar potential and the initial conditions of the scalar field evolution.

For the background cosmological sector we adopt broad flat priors on the parameters $H_0$, $\Omega_{\rm m0}$, and $\Omega_{\rm b0}$. In particular we impose the cut $40 < H_0 < 100$ on the Hubble constant, while the matter and baryon density parameters are sampled within the ranges $0.1 < \Omega_{\rm m0} < 0.9$ and $0.01 < \Omega_{\rm b0} < 0.09$. 

We also include a Big Bang Nucleosynthesis (BBN) constraint on the physical baryon density $\omega_{\rm b0}=\Omega_{\rm b0}h^2$ using the measurement
\[
\omega_{\rm b0} = 0.0222 \pm 0.0005
\]
from~\cite{Schoneberg:2024ifp}. This constraint is implemented as a Gaussian likelihood term.

Since the background expansion history is obtained by solving the scalar field equation of motion, we additionally enforce consistency between the sampled value of the Hubble constant and the value obtained from the numerical evolution. To this end we introduce a Gaussian constraint on
\[
\mathcal{E}_{H_0} \equiv 1 - \frac{H_{0,\mathrm{EOM}}^2}{H_0^2},
\]
where $H_{0,\mathrm{EOM}}$ denotes the Hubble constant derived from the equations of motion. The Gaussian width is chosen to be $0.001/\sqrt{2}$, effectively enforcing the condition $H_{0,\mathrm{EOM}}\simeq H_0$.

Both Gaussian constraints are included in the likelihood used for
parameter inference, ensuring that the sampled cosmological models
satisfy the BBN measurement and the present-day Friedmann constraint.
However, their direct contributions are excluded from the
observational likelihood employed in the statistical model comparison
and dataset concordance estimators introduced in the following
subsection. In this way, the BBN measurement is treated as external
information, while the Friedmann constraint restricts the allowed
parameter space of the generalised axion-like model.

The dark energy sector follows the scalar field model described in Section~\ref{sec:gen_axion_summary}. The parameters controlling the scalar potential and the initial field value are sampled using flat priors within the ranges $10^{-4} < n < 1$, $-2 < \log_{10}(V_{\rm min}/\rho_{\phi 0}) < 0$, $-2 < \log_{10}(f/M_{\rm Pl}) < 1$, and $-4.5 < \log_{10}(\phi_i/f) < 0.5$, where the subscript $0$ denotes present-day quantities and $i$ refers to the initial value at the starting scale factor $a_i$. The field evolution is initialised at an early epoch, $a_i/a_0 = 10^{-6}$, with vanishing initial field velocity,
\[
\dot{\phi}(a_i)=0 ,
\]
which is consistent with the freezing behaviour of the scalar field during the radiation-dominated era.

The parameters and their prior ranges adopted in the analysis are summarised in Table~\ref{tab:param_priors}. The first set corresponds to the cosmological parameters common to both $\Lambda$CDM and the scalar field model, while the remaining parameters describe the scalar potential and the initial conditions of the field.

\begin{table}[t]
\centering
\renewcommand{\arraystretch}{1.3}
\setlength{\tabcolsep}{12pt}
\begin{tabular}{c c c}
\hline\hline
Parameter & Min. & Max. \\
\hline\hline
$H_0$  & $40$ & $100$ \\
$\Omega_{\rm m0}$  & $0.1$ & $0.9$ \\
$\Omega_{\rm b0}$  & $0.01$ & $0.09$ \\
\hline
$n$ & $10^{-4}$ & $1$ \\
$\log_{10}(V_{\rm min}/\rho_{\phi 0})$ & $-2$ & $0$ \\
$\log_{10}(f/M_{\rm Pl})$ & $-2$ & $1$ \\
$\log_{10}(\phi_i/f)$ & $-4.5$ & $0.5$ \\
\hline\hline
\end{tabular}
\caption{\justifying{Prior ranges adopted for the cosmological and scalar field parameters of the generalised axion-like model. Flat priors are used in all cases.}}
\label{tab:param_priors}
\end{table}

For each sampled parameter set we solve the scalar field equation of motion from the initial scale factor $a_i$ to the present-time. The resulting background evolution is then used to compute the theoretical predictions entering the likelihood functions described in the previous subsection.

\subsection{Model comparison and dataset concordance}
\label{sec:model-comparison}

To evaluate the observational performance of the generalised axion-like
quintessence model relative to $\Lambda$CDM, we employ a collection of
statistical estimators that account for both the quality of the fit and
the complexity of the cosmological model. These diagnostics are particularly
useful when comparing models with different numbers of free parameters,
since additional degrees of freedom may improve the maximum likelihood
without necessarily providing a statistically preferred description
of the data.

Throughout the following analysis, we distinguish between the
observational likelihood associated with a given dataset combination
and the auxiliary Gaussian constraints described in
Section~\ref{sec:prior}. The latter are retained when constructing
the posterior distribution and determining the best-fitting
parameters, but their direct contributions are excluded from the
likelihood entering the statistical estimators.

Denoting the observational dataset by $\mathcal{D}$ and the auxiliary
constraints by $\mathcal{A}$, the total likelihood can be written as
\begin{equation}
P(\mathcal{D},\mathcal{A}|\boldsymbol{\theta}_M,M)
=
P(\mathcal{D}|\boldsymbol{\theta}_M,M)
P(\mathcal{A}|\boldsymbol{\theta}_M,M).
\end{equation}
For the generalised axion-like model, the auxiliary factor contains
both the BBN and Friedmann consistency likelihoods, whereas for
$\Lambda$CDM only the BBN constraint is required.

Accordingly, the observational chi-squared employed in our
statistical diagnostics is obtained by removing the auxiliary
contributions from the total chi-squared,
\begin{equation}
\chi^2_{\rm obs}
=
\chi^2_{\rm tot}
-
\chi^2_{\rm BBN}
-
\chi^2_{\mathcal{E}_{H_0}},
\label{eq:chi2_observational}
\end{equation}
where the last term is absent for $\Lambda$CDM. The posterior
distribution nevertheless retains the information provided by
the auxiliary constraints. All posterior-based estimators are
therefore evaluated conditionally on $\mathcal{A}$.

We consider estimators constructed from the maximum likelihood, namely
the corrected Akaike information criterion (${\rm AIC}_{\rm C}$) and the
Bayesian information criterion (BIC), together with posterior-based
quantities such as the Bayesian evidence, the deviance information
criterion (DIC), and the widely applicable information criterion (WAIC).
These complementary approaches allow us to examine the preference
between the generalised axion-like model and $\Lambda$CDM from different
statistical perspectives.

The maximum-likelihood estimators are evaluated using the observational
likelihood at the best-fitting parameter values obtained with the
\texttt{minimize} sampler implemented in \texttt{Cobaya}
\cite{Cartis:2018xum,Cartis:2018jxl,Powell}. The auxiliary Gaussian
constraints are retained during minimisation, while their direct
contributions are excluded from the likelihood entering the
information criteria.

The corrected Akaike information criterion is defined as
\cite{Akaike:1974vps,Liddle:2007fy}
\begin{equation}
{\rm AIC}_{\rm C}
=
-2\ln P(\mathcal{D}|\boldsymbol{\theta}_{M,{\rm bf}},M)
+
2k
+
\frac{2k(k+1)}{N-k-1},
\label{eq:aicc}
\end{equation}
where $\boldsymbol{\theta}_{M,{\rm bf}}$ denotes the best-fitting
parameter vector obtained under the auxiliary constraints, $k$
is the number of independent free parameters of the cosmological
model \footnote{Although the generalised axion-like quintessence
model introduces four additional parameters relative to background
$\Lambda$CDM, one parameter direction is effectively fixed by the
present-day Friedmann constraint, $\mathcal{E}_{H_0}=0$.},
and $N$ is the total number of observational data points in the
corresponding dataset combination. The final term corrects for
the finite size of the dataset and becomes negligible when
$N\gg k^2$, recovering the standard Akaike information criterion,
\begin{equation}
{\rm AIC}
=
-2\ln P(\mathcal{D}|\boldsymbol{\theta}_{M,{\rm bf}},M)
+
2k.
\end{equation}
In this way, ${\rm AIC}_{\rm C}$ penalises additional parameters while
retaining the information provided by the best-fitting solution.

The Bayesian information criterion is given by
\cite{Schwarz:1978tpv}
\begin{equation}
{\rm BIC}
=
-2\ln P(\mathcal{D}|\boldsymbol{\theta}_{M,{\rm bf}},M)
+
k\ln N.
\label{eq:bic}
\end{equation}
Although both criteria depend on the maximum likelihood, their
complexity penalties differ. In particular, the BIC penalty increases
logarithmically with the number of data points and generally becomes
more restrictive than that of ${\rm AIC}_{\rm C}$ for sufficiently
large datasets.

We complement these maximum-likelihood criteria with estimators
that depend on the posterior distribution. 
For any function of the model parameters, we define the posterior
average conditional on the auxiliary constraints as
\begin{equation}
\left\langle \cdots \right\rangle_{M|\mathcal{D},\mathcal{A}}
\equiv
\int
(\cdots)\,
P(\boldsymbol{\theta}_M|
\mathcal{D},\mathcal{A},M)
\,d\boldsymbol{\theta}_M.
\end{equation}
The posterior distribution appearing in this expression is obtained
from the complete likelihood, including the auxiliary constraints,
whereas the log-likelihood functions entering the statistical
estimators contain only the observational contributions. Using this notation, we define the posterior expectation value of
the log-likelihood as
\begin{equation}
F(\mathcal{D}|\mathcal{A},M)
\equiv
\left\langle
\ln P(\mathcal{D}|\boldsymbol{\theta}_M,M)
\right\rangle_{M|\mathcal{D},\mathcal{A}}.
\label{eq:F_def}
\end{equation}

Another useful posterior quantity is the Bayesian model dimensionality
(BMD) \cite{Handley:2019pqx}, defined as
\begin{align}
{\rm BMD}(\mathcal{D}|\mathcal{A},M)
&=
2\left\langle
\left[
\ln P(\mathcal{D}|\boldsymbol{\theta}_M,M)
\right]^2
\right\rangle_{M|\mathcal{D},\mathcal{A}}
\nonumber\\
&\quad
-
2\left\langle
\ln P(\mathcal{D}|\boldsymbol{\theta}_M,M)
\right\rangle_{M|\mathcal{D},\mathcal{A}}^2.
\label{eq:BMD_def}
\end{align}
This estimator characterises the effective dimensionality of the
parameter space constrained by the observations, conditional on
the auxiliary information. Unlike the nominal number of free
parameters, the BMD accounts for the information contained in
the posterior and can therefore be smaller when some parameter
combinations remain poorly determined by the data.

We next consider the Bayesian evidence, which measures the
statistical support for a cosmological model after integrating
the likelihood over its prior parameter space. Since the auxiliary
Gaussian constraints are included in the parameter inference,
we work with the Bayesian evidence conditional on these constraints.

Let $P(\boldsymbol{\theta}_M|M)$ denote the parameter prior, and let $\mathcal{A}$ represent the auxiliary information
introduced in Section~\ref{sec:prior}. We define the corresponding
normalised effective prior as
\begin{equation}
P(\boldsymbol{\theta}_M|\mathcal{A},M)
\equiv
\frac{
P(\mathcal{A}|\boldsymbol{\theta}_M,M)
P(\boldsymbol{\theta}_M|M)
}{
V_{\mathcal{A},M}
},
\label{eq:effective_prior}
\end{equation}
where
\begin{equation}
V_{\mathcal{A},M}
=
\int_{\Pi_M}
P(\mathcal{A}|\boldsymbol{\theta}_M,M)
P(\boldsymbol{\theta}_M|M)
\,d\boldsymbol{\theta}_M.
\label{eq:aux_prior_volume}
\end{equation}

The conditional Bayesian evidence is then defined as
\begin{equation}
\begin{aligned}
P(\mathcal{D}|\mathcal{A},M)
\equiv{}&
\int_{\Pi_M}
P(\mathcal{D}|\boldsymbol{\theta}_M,M)
\\
&\times
P(\boldsymbol{\theta}_M|\mathcal{A},M)
\,d\boldsymbol{\theta}_M.
\end{aligned}
\label{eq:evidence_def}
\end{equation}
Equivalently, it can be expressed as the ratio
\begin{equation}
P(\mathcal{D}|\mathcal{A},M)
=
\frac{
P(\mathcal{D},\mathcal{A}|M)
}{
P(\mathcal{A}|M)
},
\label{eq:conditional_evidence_ratio}
\end{equation}
where $P(\mathcal{D},\mathcal{A}|M)$ denotes the joint evidence
obtained by integrating the product of the observational and
auxiliary likelihoods over the parameter prior, while
$P(\mathcal{A}|M)=V_{\mathcal{A},M}$ is the evidence associated
with the auxiliary constraints alone.

For consistency with the information criteria, we introduce
the rescaled logarithmic conditional evidence
\begin{equation}
\ln {\rm B}(\mathcal{D}|\mathcal{A},M)
\equiv
2\ln P(\mathcal{D}|\mathcal{A},M),
\label{eq:lnB_def}
\end{equation}
so that
\begin{equation}
-\ln {\rm B}(\mathcal{D}|\mathcal{A},M)
=
-2\ln P(\mathcal{D}|\mathcal{A},M).
\label{eq:minus_logB}
\end{equation}
Using Bayes' theorem, the inverse conditional evidence can
be expressed as a posterior average,
\begin{equation}
\begin{aligned}
&P(\mathcal{D}|\mathcal{A},M)^{-1}
\\
&\quad =
\left\langle
\left[
P(\mathcal{D}|\boldsymbol{\theta}_M,M)
\right]^{-1}
\right\rangle_{M|\mathcal{D},\mathcal{A}}.
\end{aligned}
\label{eq:conditional_evidence_inverse}
\end{equation}
Consequently,
\begin{equation}
\begin{aligned}
&-\ln {\rm B}(\mathcal{D}|\mathcal{A},M)
\\
&\quad =
2\ln
\left\langle
\left[
P(\mathcal{D}|\boldsymbol{\theta}_M,M)
\right]^{-1}
\right\rangle_{M|\mathcal{D},\mathcal{A}}.
\end{aligned}
\label{eq:minus_logB_harmonic}
\end{equation}

We also employ the deviance information criterion, which combines
the likelihood evaluated at the maximum-a-posteriori point with the
posterior-averaged log-likelihood. Following~\cite{Celeux},
we define
\begin{equation}
\begin{aligned}
{\rm DIC}(\mathcal{D}|\mathcal{A},M)
=
2\ln P(\mathcal{D}|
\boldsymbol{\theta}_{M,{\rm map}},M)
-
4F(\mathcal{D}|\mathcal{A},M).
\end{aligned}
\label{eq:DIC_def}
\end{equation}
where $\boldsymbol{\theta}_{M,{\rm map}}$ denotes the parameter
vector that maximises the posterior probability, including the
auxiliary Gaussian constraints described in Section~\ref{sec:prior}.
This point is determined using the \texttt{minimize} sampler
implemented in \texttt{Cobaya}
\cite{Cartis:2018xum,Cartis:2018jxl,Powell}. However, the likelihood
entering the DIC is evaluated using only the observational
contributions, consistently with the conditional framework
adopted throughout this analysis. The DIC incorporates
the effective complexity of the posterior distribution, rather than
penalising the model solely according to its nominal number of
parameters.

Finally, following the posterior-based convention adopted here,
we define the widely applicable information criterion as
\cite{Watanabe:2012awa}
\begin{equation}
{\rm WAIC}(\mathcal{D}|\mathcal{A},M)
=
-2F(\mathcal{D}|\mathcal{A},M)
+
{\rm BMD}(\mathcal{D}|\mathcal{A},M).
\label{eq:WAIC_def}
\end{equation}
This estimator includes both the average log-likelihood and the
posterior fluctuations of the log-likelihood, thereby incorporating
information about the distribution of statistically allowed
parameter values.

To compare the generalised axion-like quintessence model with the
reference $\Lambda$CDM cosmology, we introduce
\begin{equation}
\Delta{\rm IC}
\equiv
{\rm IC}_{\rm axion}
-
{\rm IC}_{\Lambda{\rm CDM}},
\label{eq:Delta_IC}
\end{equation}
where
\begin{equation}
{\rm IC}
\in
\left\{
{\rm AIC}_{\rm C},
{\rm BIC},
-\ln{\rm B},
{\rm DIC},
{\rm WAIC}
\right\}.
\end{equation}
According to this convention, $\Delta{\rm IC}>0$ indicates a
preference for $\Lambda$CDM, whereas $\Delta{\rm IC}<0$ favours the
generalised axion-like quintessence model. We interpret the
differences according to the Jeffreys'-scale convention summarised
in Table~\ref{tab:jeffreys_scale}.

\begin{table}[t]
\centering
\renewcommand{\arraystretch}{1.3}
\setlength{\tabcolsep}{10pt}
\begin{tabular}{c c}
\hline\hline
$\Delta{\rm IC}$ & Interpretation \\
\hline\hline
$>10$ & Strongly disfavoured / tensioned \\
$5\sim10$ & Moderately disfavoured / tensioned \\
$2\sim5$ & Weakly disfavoured / tensioned \\
$-2\sim2$ & Inconclusive \\
$-5\sim-2$ & Weakly favoured / aligned \\
$-10\sim-5$ & Moderately favoured / aligned \\
$<-10$ & Strongly favoured / aligned \\
\hline\hline
\end{tabular}
\caption{\justifying{Interpretation of information criterion differences
and dataset concordance diagnostics following a Jeffreys'-scale
convention. Positive values indicate that the generalised axion-like
quintessence model is disfavoured relative to $\Lambda$CDM, or that
the datasets exhibit more tension. Negative values correspond to model
preference or improved dataset agreement, whereas values close to
zero are inconclusive.}}
\label{tab:jeffreys_scale}
\end{table}

Beyond the relative statistical preference between cosmological
models, we investigate whether the generalised axion-like quintessence
scenario changes the consistency between different observational
datasets. This question is particularly relevant in the context of
the Hubble tension, since the distance-ladder measurements and the
combination of CMB and BAO observations favour different values of
the present expansion rate.

We therefore examine the concordance between the CMB+BAO dataset
and the two alternative late-time combinations considered in this
work, namely PPS and CC+GRB+Pan. The latter two are analysed
separately and are never included simultaneously in the same
dataset combination. To characterise their compatibility, we
consider the Bayesian evidence ratio, the goodness-of-fit statistic,
and suspiciousness.

For two observational datasets $\mathcal{D}_1$ and
$\mathcal{D}_2$, we define the Bayesian concordance ratio
conditionally on the auxiliary constraints. The corresponding
rescaled logarithmic statistic is
\begin{align}
-\ln {\rm R}
(\mathcal{D}_1,\mathcal{D}_2|\mathcal{A},M)
\equiv&
-\ln {\rm B}
(\mathcal{D}_1\mathcal{D}_2|\mathcal{A},M)
\nonumber\\
&+
\ln {\rm B}(\mathcal{D}_1|\mathcal{A},M)
\nonumber\\
&+
\ln {\rm B}(\mathcal{D}_2|\mathcal{A},M).
\label{eq:lnR_def}
\end{align}
Here, the same auxiliary information $\mathcal{A}$ is imposed
in the separate and joint dataset analyses. The resulting
statistic measures the concordance between the observational
datasets after conditioning on the BBN constraint and, for the
generalised axion-like model, on the present-day Friedmann
consistency condition.

To assess whether the generalised axion-like scenario modifies the
dataset agreement relative to $\Lambda$CDM, we define
\begin{equation}
\begin{aligned}
\Delta(-\ln {\rm R})
={}&
[-\ln {\rm R}]_{\rm axion}
\\
&-
[-\ln {\rm R}]_{\Lambda{\rm CDM}},
\end{aligned}
\label{eq:Delta_lnR}
\end{equation}
where both terms are evaluated using their corresponding
conditional Bayesian evidences. A positive value indicates that the generalised axion-like model
increases the discrepancy between the datasets compared with
$\Lambda$CDM, whereas a negative value indicates an improvement
in their concordance. The magnitude of this difference is interpreted
using Table~\ref{tab:jeffreys_scale}.

We further consider the goodness-of-fit concordance statistic
\cite{Raveri:2019gdp,DES:2020hen}, which quantifies the loss in
best-fit quality when two datasets are described by a common set
of cosmological parameters. It is defined as
\begin{align}
{\rm GoF}(\mathcal{D}_1,\mathcal{D}_2|M)
\equiv&
-\ln P(\mathcal{D}_1\mathcal{D}_2|
\boldsymbol{\theta}_{M,{\rm map}}^{12},M)
\nonumber\\
&+
\ln P(\mathcal{D}_1|
\boldsymbol{\theta}_{M,{\rm map}}^{1},M)
\nonumber\\
&+
\ln P(\mathcal{D}_2|
\boldsymbol{\theta}_{M,{\rm map}}^{2},M),
\label{eq:GoF_def}
\end{align}
where $\boldsymbol{\theta}_{M,{\rm map}}^{12}$ denotes the
maximum-a-posteriori parameter vector obtained from the combined
dataset, while $\boldsymbol{\theta}_{M,{\rm map}}^{1}$ and
$\boldsymbol{\theta}_{M,{\rm map}}^{2}$ correspond to the separate
analyses \footnote{The maximum-a-posteriori parameter values are determined using
the posterior distribution that includes the auxiliary constraints,
while the likelihoods appearing in the goodness-of-fit statistic
contain only the corresponding observational contributions.}. This statistic measures the extent to which combining
the datasets forces a deterioration of the best-fitting solutions.

Another concordance diagnostic is suspiciousness, which is
constructed from the posterior-averaged log-likelihoods rather
than their values at a single point in parameter space
\cite{DES:2020hen}. It is given by
\begin{align}
{\rm S}(\mathcal{D}_1,\mathcal{D}_2|\mathcal{A},M)
\equiv{}&
-F(\mathcal{D}_1\mathcal{D}_2|\mathcal{A},M)
\nonumber\\
&+
F(\mathcal{D}_1|\mathcal{A},M)
+
F(\mathcal{D}_2|\mathcal{A},M).
\label{eq:S_def}
\end{align}
Since suspiciousness incorporates the likelihood information
averaged over the posterior, it provides a complementary assessment
of dataset compatibility to that obtained from the goodness-of-fit
statistic.

To quantify the significance associated with these concordance
estimators, we introduce the effective Bayesian model dimensionality
of the dataset comparison, conditional on the auxiliary constraints,
\begin{align}
{\rm BMD}_{\rm eff}
={}&
-{\rm BMD}(\mathcal{D}_1\mathcal{D}_2|\mathcal{A},M)
\nonumber\\
&+
{\rm BMD}(\mathcal{D}_1|\mathcal{A},M)
\nonumber\\
&+
{\rm BMD}(\mathcal{D}_2|\mathcal{A},M).
\label{eq:BMD_eff}
\end{align}
Under the Gaussian approximation, the goodness-of-fit and
suspiciousness can be related to an effective chi-squared statistic
through
\begin{equation}
Q_{\mathcal{C}}
=
{\rm BMD}_{\rm eff}
+
2\mathcal{C},
\label{eq:Q_concordance}
\end{equation}
where $\mathcal{C}$ denotes either ${\rm GoF}$ or ${\rm S}$.
The quantity $Q_{\mathcal{C}}$ approximately follows a chi-squared
distribution with ${\rm BMD}_{\rm eff}$ effective degrees of freedom
\cite{Handley:2019wlz,Lemos:2019txn}.

The corresponding tension can be expressed as an equivalent
one-dimensional Gaussian significance,
\begin{equation}
\sigma_{\mathcal{C}}
=
{\rm CDF}_1^{-1}
\left[
{\rm CDF}_{{\rm BMD}_{\rm eff}}
\left(
\sqrt{Q_{\mathcal{C}}}
\right)
\right],
\label{eq:sigma_concordance}
\end{equation}
where ${\rm CDF}_{d}(r)$ is the radial cumulative distribution
function of a $d$-dimensional standard Gaussian,
\begin{equation}
{\rm CDF}_{d}(r)
=
\frac{1}{2^{d/2}\Gamma(d/2)}
\int_0^{r^2}
x^{d/2-1}e^{-x/2}\,dx,
\end{equation}
with
\begin{equation}
\Gamma(x)
=
\int_0^\infty
t^{x-1}e^{-t}\,dt
\end{equation}
denoting the Gamma function.

The information criteria and concordance estimators provide
different but complementary tests of the generalised axion-like
quintessence model. While the former establish whether the
additional scalar field parameters lead to a statistically
justified improvement over $\Lambda$CDM, the latter determine
whether the model changes the level of consistency between
cosmological observations obtained at different redshifts.

In practice, we evaluate these estimators using the independent
MCMC chains generated for each cosmological model and dataset
combination. Denoting the number of available chains by $N_{\rm c}$,
which varies between two and four in our analysis, we assess the
stability of the numerical results by comparing the values
obtained from the different chains. This procedure is applied
to ${\rm AIC}_{\rm C}$, BIC, Bayesian evidence, DIC, WAIC,
Bayes ratio, goodness-of-fit, and suspiciousness.

For a generic estimator $\mathcal{O}$, with $\mathcal{O}_i$
representing the value obtained from the $i$th chain, we define
the inter-chain mean as
\begin{equation}
\overline{\mathcal{O}}
=
\frac{1}{N_{\rm c}}
\sum_{i=1}^{N_{\rm c}}
\mathcal{O}_i,
\end{equation}
and the corresponding standard deviation as
\begin{equation}
s_{\mathcal{O}}
=
\sqrt{
\frac{1}{N_{\rm c}-1}
\sum_{i=1}^{N_{\rm c}}
\left(
\mathcal{O}_i-
\overline{\mathcal{O}}
\right)^2
}.
\end{equation}
All chains contribute with equal weight to these estimates.
Unless explicitly stated otherwise, the statistical estimators
are reported in the form
$\overline{\mathcal{O}}\pm s_{\mathcal{O}}$.

\section{Observational constraints of our model} \label{fits}

\begin{table*}[ht!]
\centering
\scriptsize
\setlength{\tabcolsep}{3pt}

\resizebox{\textwidth}{!}{%
\begin{tabular}{lcccccc}
\hline
\hline
\addlinespace[2pt]
Parameter
& BAO
& CMB+BAO
& PPS
& CC+GRB+Pan
& CMB+BAO+PPS
& CMB+BAO+CC+GRB+Pan \\
\addlinespace[2pt]
\hline
\hline
\addlinespace[2pt]

\multicolumn{1}{c}{}
& \multicolumn{6}{c}{\textbf{$\Lambda$CDM}} \\
\addlinespace[2pt]
\hline
\addlinespace[2pt]

$H_0$
& $68.62 \pm 0.51$ & $68.40 \pm 0.30$ & $73.6 \pm 1.0$ & $68.9 \pm 1.9$ & $68.71 \pm 0.27$ & $68.42 \pm 0.28$ \\

$\Omega_{\rm m0}$
& $0.2975 \pm 0.0089$ & $0.3011 \pm 0.0038$ & $0.332 \pm 0.018$ & $0.300 \pm 0.020$ & $0.2974 \pm 0.0034$ & $0.3010 \pm 0.0035$ \\

$\Omega_{\rm b0}$
& $0.04713 \pm 0.00073$ & $0.04813 \pm 0.00034$ & $0.0411 \pm 0.0015$ & $0.0469^{+0.0024}_{-0.0030}$ & $0.04789 \pm 0.00032$ & $0.04813 \pm 0.00032$ \\

\addlinespace[2pt]
\hline
\addlinespace[2pt]

AIC$_{\rm C}$
& $18.9532 \pm 0.0023$ & $24.2468 \pm 0.0056$ & $1458.0370 \pm 0.0065$ & $103.63971 \pm 0.00052$ & $1506.099 \pm 0.028$ & $119.9878 \pm 0.0074$ \\

BIC
& $17.9814 \pm 0.0023$ & $24.5646 \pm 0.0056$ & $1474.3397 \pm 0.0065$ & $112.54919 \pm 0.00052$ & $1522.430 \pm 0.028$ & $129.2133 \pm 0.0074$ \\

DIC
& $14.36 \pm 0.20$ & $22.20 \pm 0.15$ & $1455.963 \pm 0.059$ & $101.58 \pm 0.22$ & $1505.92 \pm 0.19$ & $119.64 \pm 0.23$ \\

WAIC
& $14.59 \pm 0.53$ & $22.21 \pm 0.44$ & $1455.94 \pm 0.10$ & $101.73 \pm 0.36$ & $1505.95 \pm 0.36$ & $119.55 \pm 0.22$ \\

$-\ln{\rm B}$
& $16.46 \pm 0.64$ & $22.37 \pm 0.68$ & $1456.29 \pm 0.51$ & $102.36 \pm 0.89$ & $1506.49 \pm 0.44$ & $119.34 \pm 0.12$ \\

\addlinespace[2pt]
\hline
\addlinespace[2pt]

\multicolumn{1}{c}{}
& \multicolumn{6}{c}{\textbf{Generalised axion-like}} \\
\addlinespace[2pt]
\hline
\addlinespace[2pt]

$H_0$
& $66.9^{+2.2}_{-1.1}$ & $68.23 \pm 0.39$ & $73.6 \pm 1.0$ & $68.9 \pm 1.8$ & $68.70 \pm 0.28$ & $68.37 \pm 0.31$ \\

$\Omega_{\rm m0}$
& $0.294^{+0.012}_{-0.0098}$ & $0.3025 \pm 0.0043$ & $0.285^{+0.068}_{-0.032}$ & $0.286^{+0.034}_{-0.020}$ & $0.2973 \pm 0.0035$ & $0.3011^{+0.0036}_{-0.0040}$ \\

$\Omega_{\rm b0}$
& $0.0497^{+0.0016}_{-0.0033}$ & $0.04840^{+0.00041}_{-0.00057}$ & $0.0411 \pm 0.0015$ & $0.0469 \pm 0.0026$ & $0.04792 \pm 0.00033$ & $0.04820^{+0.00033}_{-0.00040}$ \\

$n$
& $<0.519$ & $<0.606$ & -- & $<0.612$ & -- & -- \\

$\log_{10}(V_{\min}/\rho_{\phi0})$
& $>-0.163$ & $>-0.0239$ & $>-0.189$ & $>-0.0267$ & $>-0.000573$ & $>-0.000863$ \\

$\log_{10}(f/M_{\rm Pl})$
& $>-0.722$ & -- & $-0.34^{+0.99}_{-0.70}$ & $-0.66^{+0.52}_{-1.1}$ & $<-1.45$ & $<-1.21$ \\

$\log_{10}(\phi_i/f)$
& -- & $>-1.89$ & -- & -- & $>-2.73$ & $>-2.49$ \\

\addlinespace[2pt]
\hline
\addlinespace[2pt]

AIC$_{\rm C}$
& $34.643 \pm 0.097$ & $37.5909 \pm 0.0084$ & $1463.7860 \pm 0.0086$ & $109.867 \pm 0.044$ & $1512.1870 \pm 0.0019$ & $126.4002 \pm 0.0036$ \\

BIC
& $24.033 \pm 0.097$ & $32.8931 \pm 0.0084$ & $1496.3702 \pm 0.0086$ & $127.431 \pm 0.044$ & $1544.8279 \pm 0.0019$ & $144.6223 \pm 0.0036$ \\

DIC
& $15.30 \pm 0.43$ & $23.391 \pm 0.093$ & $1456.34 \pm 0.16$ & $102.35 \pm 0.13$ & $1507.588 \pm 0.074$ & $121.05 \pm 0.31$ \\

WAIC
& $14.24 \pm 0.46$ & $23.17 \pm 0.16$ & $1456.030 \pm 0.019$ & $102.2719 \pm 0.0083$ & $1507.382 \pm 0.069$ & $120.85 \pm 0.26$ \\

$-\ln{\rm B}$
& $15.09 \pm 0.73$ & $24.16 \pm 0.37$ & $1456.9 \pm 1.4$ & $104.4 \pm 2.5$ & $1508.263 \pm 0.080$ & $121.27 \pm 0.62$ \\

\addlinespace[2pt]
\hline
\addlinespace[2pt]

\multicolumn{1}{c}{}
& \multicolumn{6}{c}{
\textbf{Differences relative to $\Lambda$CDM}
} \\
\addlinespace[2pt]
\hline
\addlinespace[2pt]

$\Delta{\rm AIC}_{\rm C}$
& $15.690 \pm 0.097$ & $13.344 \pm 0.010$ & $5.749 \pm 0.011$ & $6.227 \pm 0.044$ & $6.088 \pm 0.028$ & $6.4124 \pm 0.0082$ \\

$\Delta{\rm BIC}$
& $6.052 \pm 0.097$ & $8.329 \pm 0.010$ & $22.031 \pm 0.011$ & $14.882 \pm 0.044$ & $22.398 \pm 0.028$ & $15.4090 \pm 0.0082$ \\

$\Delta{\rm DIC}$
& $0.94 \pm 0.47$ & $1.19 \pm 0.18$ & $0.38 \pm 0.17$ & $0.77 \pm 0.26$ & $1.67 \pm 0.20$ & $1.41 \pm 0.39$ \\

$\Delta{\rm WAIC}$
& $-0.35 \pm 0.70$ & $0.96 \pm 0.47$ & $0.09 \pm 0.10$ & $0.54 \pm 0.36$ & $1.43 \pm 0.37$ & $1.30 \pm 0.34$ \\

$\Delta(-\ln{\rm B})$
& $-1.37 \pm 0.97$ & $1.79 \pm 0.77$ & $0.6 \pm 1.5$ & $2.04 \pm 2.7$ & $1.77 \pm 0.45$ & $1.93 \pm 0.63$ \\

\addlinespace[2pt]
\hline
\hline
\end{tabular}%
}

\caption{\justifying{
Summary of the observational constraints obtained
for $\Lambda$CDM and the generalised axion-like
dark energy model, using the different
dataset combinations considered in this work:
BAO, CMB+BAO, PPS, CC+GRB+Pan, CMB+BAO+PPS,
and CMB+BAO+CC+GRB+Pan.
Reported parameter constraints correspond to
posterior mean values and $68\%$ credible intervals. Parameters that are unconstrained are omitted. The table also reports the model comparison
statistics $\mathrm{AIC}_{\rm C}$, $\mathrm{BIC}$,
$\mathrm{DIC}$, $\mathrm{WAIC}$, and
$-\ln{\rm B}$, together with their differences
relative to $\Lambda$CDM.
Positive values of
$\Delta\mathrm{AIC}_{\rm C}$,
$\Delta\mathrm{BIC}$,
$\Delta\mathrm{DIC}$,
$\Delta\mathrm{WAIC}$, and
$\Delta(-\ln{\rm B})$
indicate that the quintessence model is disfavoured relative
to $\Lambda$CDM, according to the convention
adopted in Section~\ref{sec:model-comparison}.
}}
\label{tab:galde_results}
\end{table*}

In this section, we present the result of MCMC according to the methodology presented in section~\ref{sec:methodology}. As mentioned previously, the dataset we consider are grouped into 2 categories: CMB-excluding datasets of BAO, PPS, CC+GRB+Pan, and CMB-including datasets of CMB+BAO, CMB+BAO+PPS, CMB+BAO+CC+GRB+Pan, with CMB representing CMB distance prior \cite{Zhai:2018vmm} rather than the full CMB power spectrum fit.

The 68\% confidence intervals of cosmological parameters and the tension probes for $\Lambda$CDM model and the axion-like dark energy model considered in Section~\ref{sec:gen_axion_summary} given various combination of datasets are shown in Tables~\ref{tab:galde_results} and \ref{tab:galde_dataset_tension}.

\begin{figure*}[ht!]
    \centering

    % Top panel: BAO vs CMB+BAO
    \begin{subfigure}{0.60\textwidth}
        \centering
        \includegraphics[width=\linewidth]{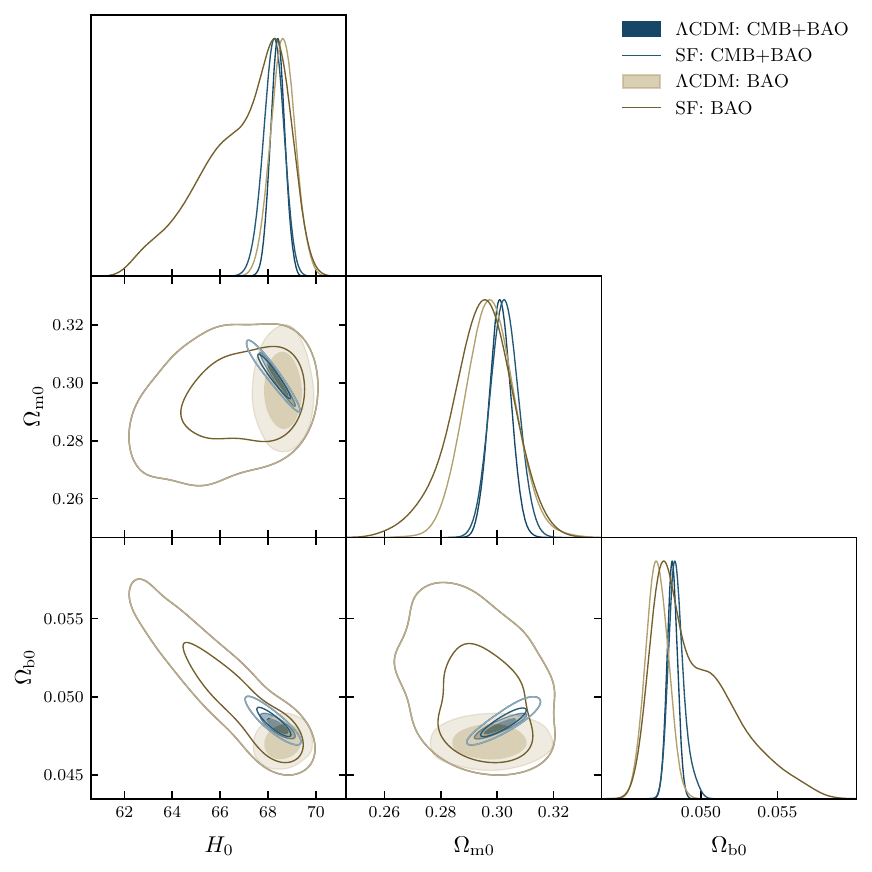}
        \caption{BAO and CMB+BAO.}
        \label{fig:galde_bao}
    \end{subfigure}

    \vspace{0.5cm}

    % Bottom-left panel: PPS vs CMB+BAO
    \begin{subfigure}{0.49\textwidth}
        \centering
        \includegraphics[width=\linewidth]{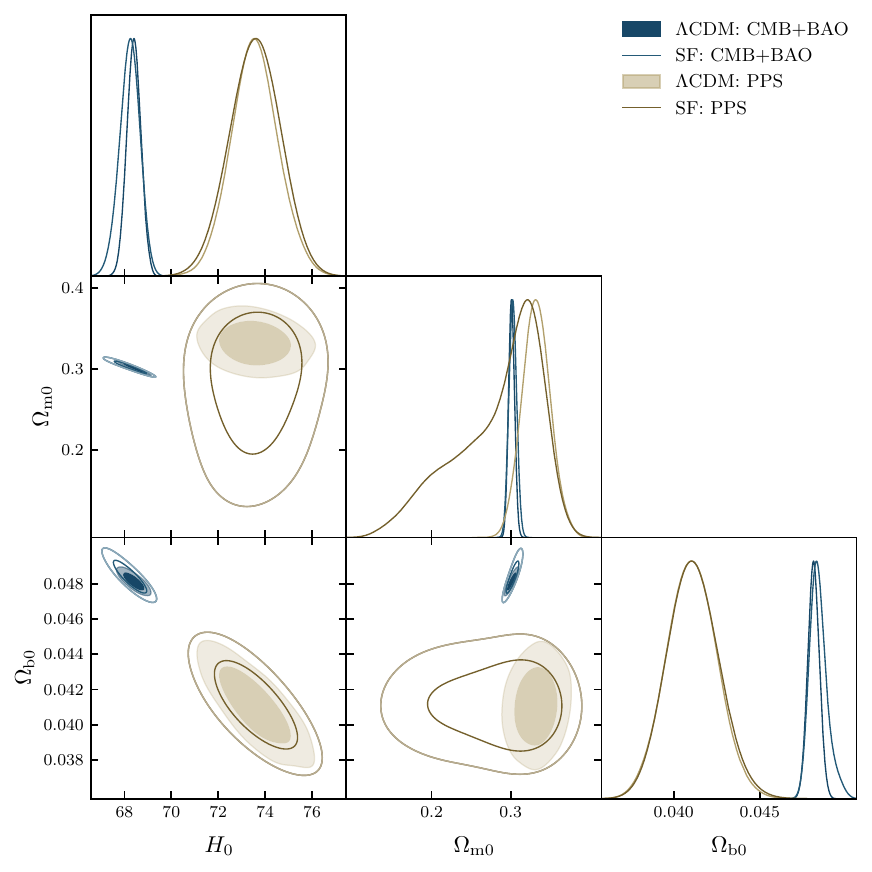}
        \caption{PPS and CMB+BAO.}
        \label{fig:galde_pps}
    \end{subfigure}
    \hfill
    % Bottom-right panel: CC+GRB+Pan vs CMB+BAO
    \begin{subfigure}{0.49\textwidth}
        \centering
        \includegraphics[width=\linewidth]{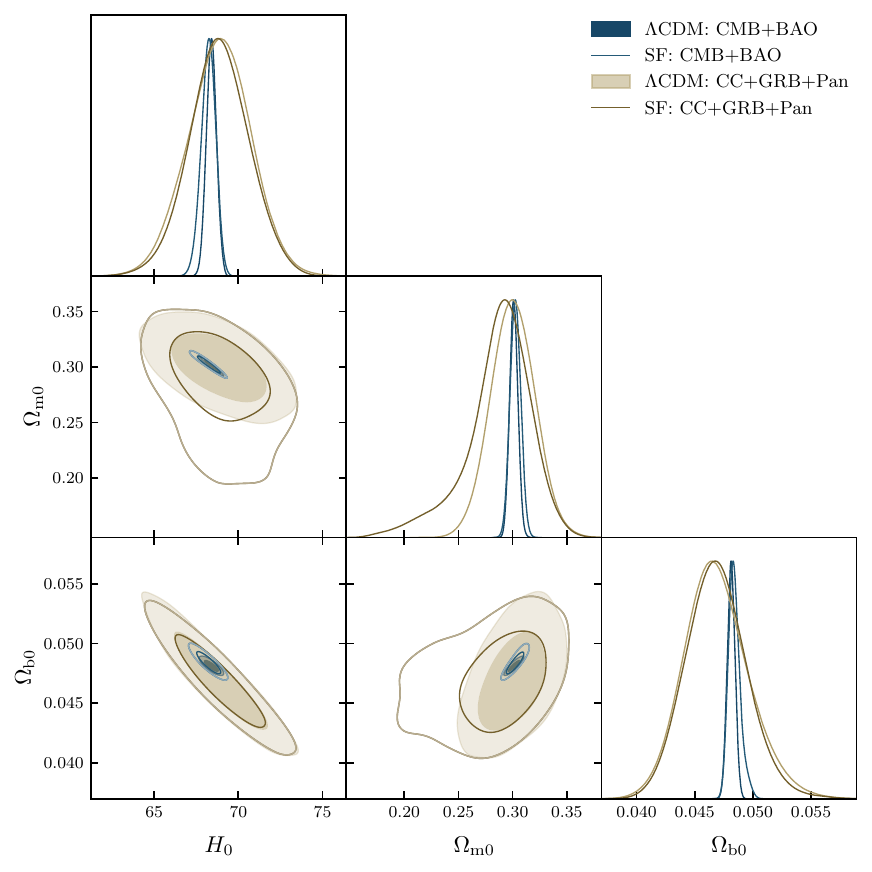}
        \caption{CC+GRB+Pan and CMB+BAO.}
        \label{fig:galde_cc}
    \end{subfigure}

    \caption{\justifying{
        Marginalised one-dimensional posterior distributions and
        two-dimensional 68\% and 95\% credible regions for
        $H_0$, $\Omega_{\mathrm{m}0}$, and $\Omega_{\mathrm{b}0}$.
        We compare the $\Lambda$CDM and scalar field models
        using CMB+BAO constraints (blue) against
        BAO-only constraints (top),
        PPS (bottom left), and
        CC+GRB+Pan (bottom right).
    }}
    \label{fig:galde_posterior_comparison}

\end{figure*}

The BAO-only constraints already illustrate the additional freedom of the
scalar field model. Whereas $\Lambda$CDM gives
$H_0=68.62\pm0.51\,{\rm km\,s^{-1}\,Mpc^{-1}}$, the axion-like model yields
$H_0=66.9^{+2.2}_{-1.1}\,{\rm km\,s^{-1}\,Mpc^{-1}}$.
The broader and asymmetric scalar field posterior is narrowed considerably
when the CMB distance prior is included. For consistency check, we verify that the background-level fit of CMB based on the distance prior approach is consistent with full perturbation analysis of \cite{Chiang:2025qxg} (c.f. CMB + BAO and CMB + BAO + PPS result in Tables~\ref{tab:galde_results} and \ref{tab:galde_dataset_tension}). As shown in Fig.~\ref{fig:galde_posterior_comparison}, for PPS dataset the axion-like dark energy model we consider tends to have a lower $\Omega_{\rm m0}$ to compensate for the temporary reduction of the deceleration parameter by the increase of dark energy equation of state to $w_\phi \sim - \left( 1 + n \right)^{-1}$ in the tracking regime. This effect on $\Omega_{\rm m0}$ is absent in the case of CMB + BAO dataset, which is well explained by the anchoring of BAO distance measure and the CMB distance prior.

\begin{figure*}[ht!]
    \centering

    % Top: BAO
    \begin{subfigure}{0.60\textwidth}
        \centering
        \includegraphics[width=\linewidth]{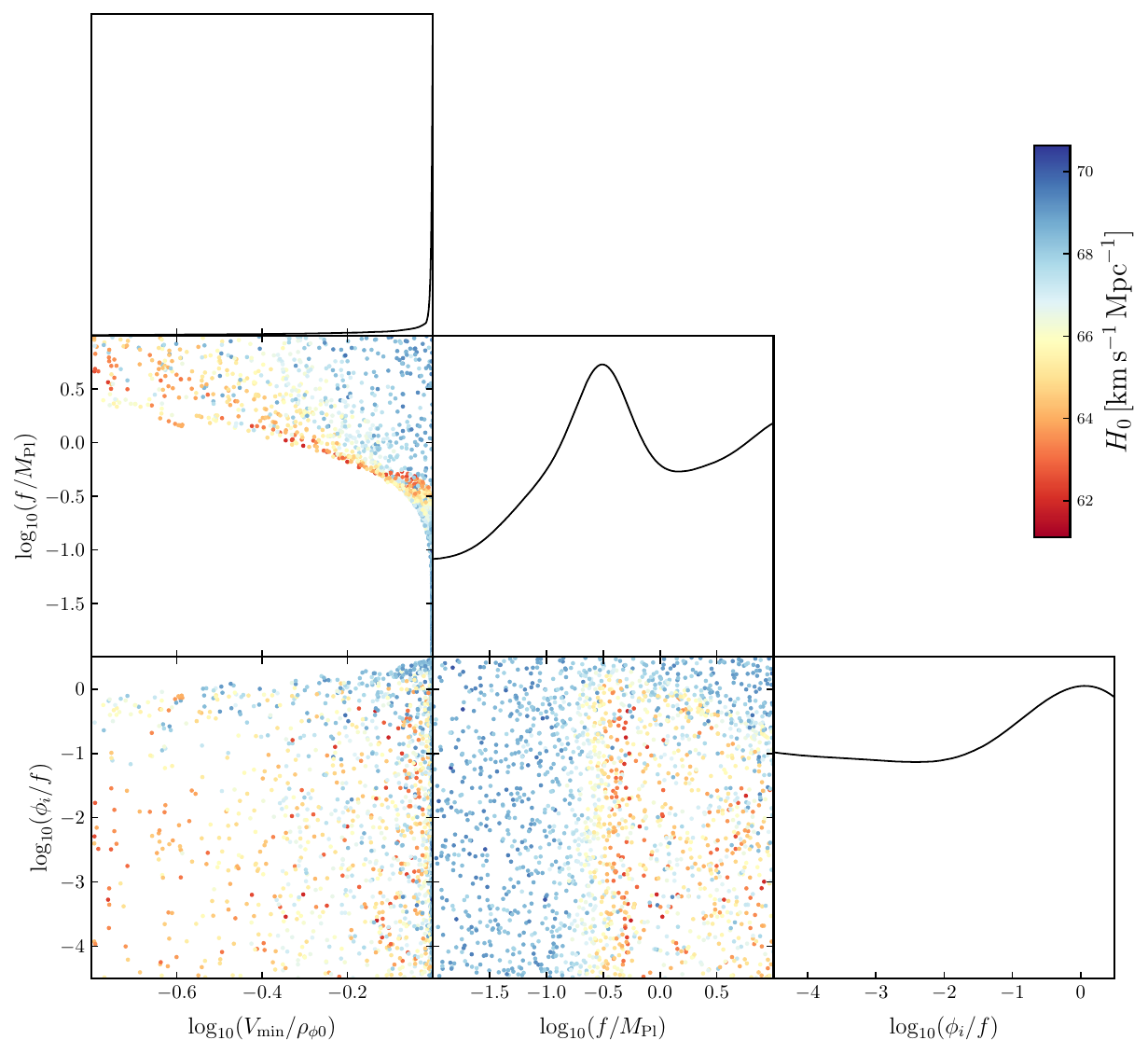}
        \caption{BAO.}
        \label{fig:galde_scatter_bao}
    \end{subfigure}

    \vspace{0.5cm}

    % Bottom left: PPS
    \begin{subfigure}{0.49\textwidth}
        \centering
        \includegraphics[width=\linewidth]{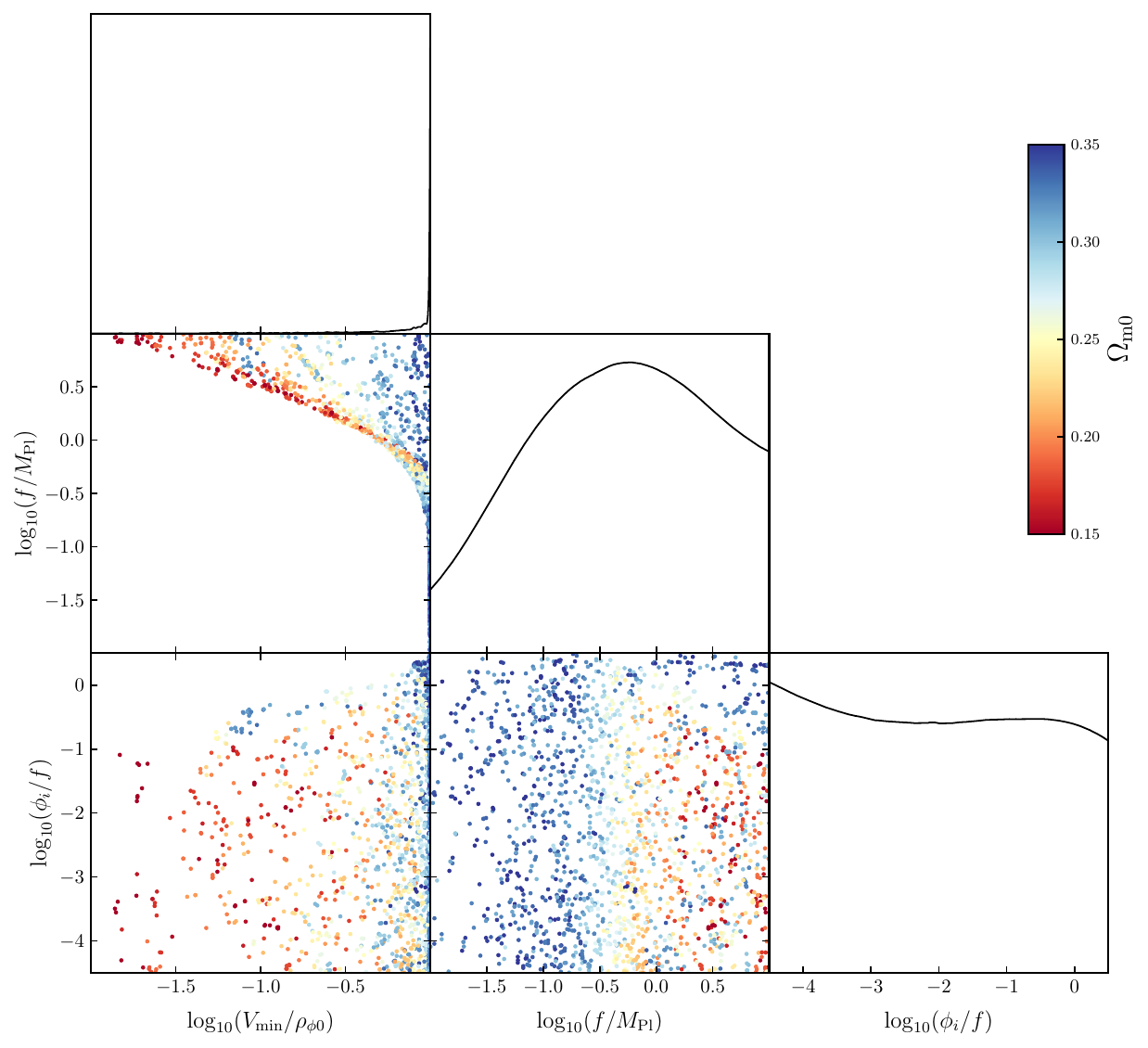}
        \caption{PPS.}
        \label{fig:galde_scatter_pps}
    \end{subfigure}
    \hfill
    % Bottom right: CC + GRB + Pantheon+
    \begin{subfigure}{0.49\textwidth}
        \centering
        \includegraphics[width=\linewidth]{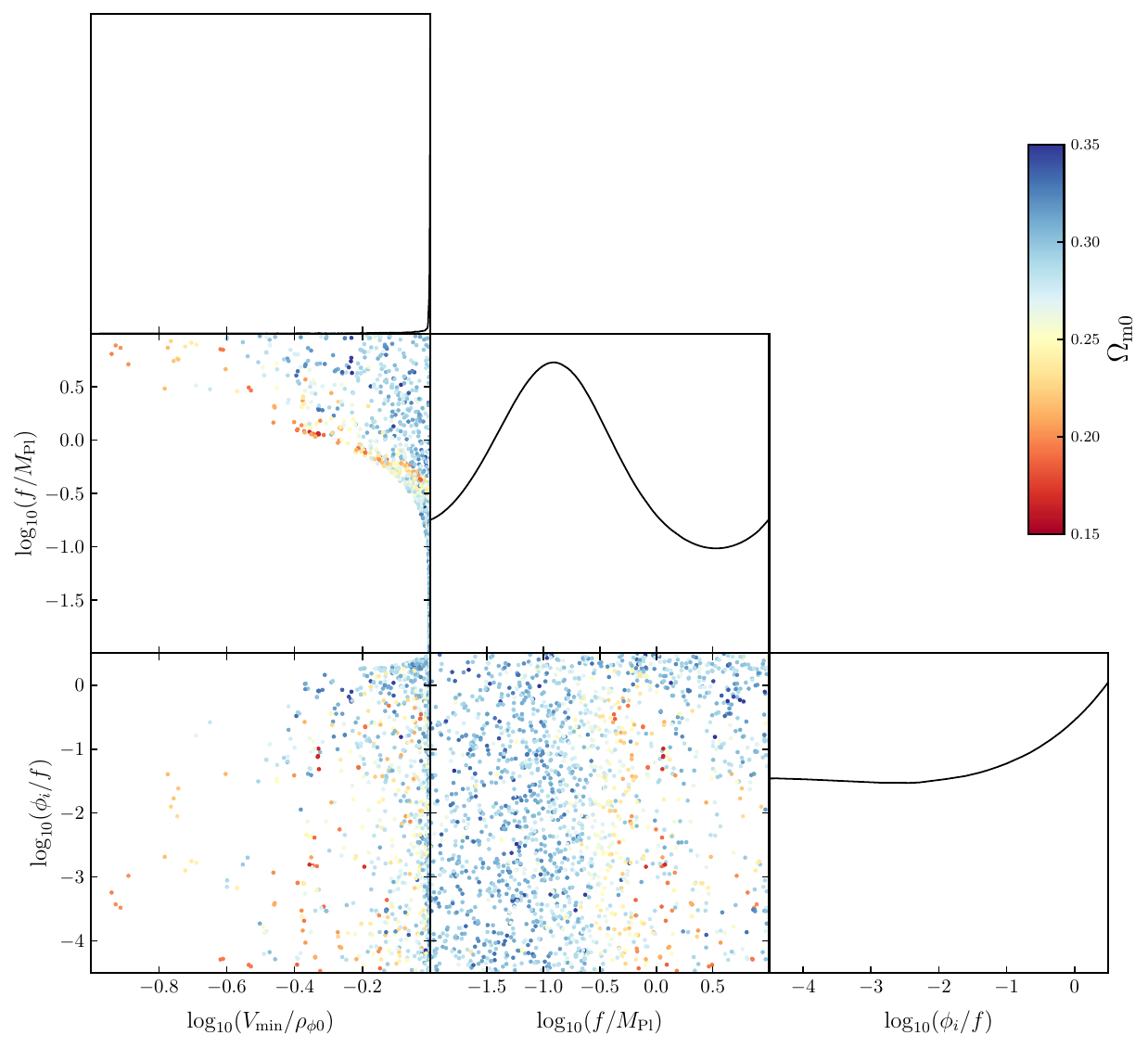}
        \caption{CC+GRB+Pan.}
        \label{fig:galde_scatter_cc}
    \end{subfigure}

    \caption{\justifying{
        Marginalised one-dimensional posterior distributions
        and two-dimensional posterior scatter plots for the
        generalised axion-like model parameters
        $\log_{10}(V_{\min}/\rho_{\phi0})$,
        $\log_{10}(f/M_{\rm Pl})$, and
        $\log_{10}(\phi_i/f)$.
        The panels correspond to BAO-only constraints (top),
        PPS (bottom left), and
        CC+GRB+Pan (bottom right).
        The scatter points are coloured according to
        $H_0$ in the BAO-only panel and
        $\Omega_{\rm m0}$ in the two lower panels,
        as indicated by their respective colour bars.
    }}
    \label{fig:galde_scalar_posterior_scatter}

\end{figure*}

We then turn our attention to the newly introduced CC + GRB + Pan. As CC provides tighter constraint on cosmic evolution at mid-to-low redshift than SNe Ia alone via direct Hubble parameter measurements, the $\Omega_{\rm m0}$ contour is clearly narrower in Fig.~\ref{fig:galde_posterior_comparison}. This is further confirmed by Fig.~\ref{fig:galde_scalar_posterior_scatter} where the parameter space leading to a lower $\Omega_{\rm m0}$ is clearly disfavoured. Effectively, CC + GRB + Pan prefers $\Lambda$CDM-esque branch. The combined dataset result of CMB + BAO + CC + GRB + Pan shows similar non-preference away from $\Lambda$CDM, with strong consistency between CMB + BAO and CC + GRB + Pan. As such, the added parameters serves no purpose and only incurs strong debt on IC probes, as shown in Table~\ref{tab:galde_results}.

We would like to emphasise that while eventually the scalar field behaves like a cosmological constant at late-time, the tracking still happens in earlier matter-dominated era. The scalar dark energy model we consider therefore serves as a natural extension to $\Lambda$CDM. With additional degrees of freedom, the tracking may serves as a window to more complicated phenomenon \cite{Calderon:2026vgr}.

The scalar field parameter bounds provide a complementary view of this
preference. Table~\ref{tab:galde_results} shows that the four additional
parameters are not all independently determined: several posteriors admit
only one-sided bounds or remain unconstrained. In particular, the
CMB-containing combinations require
$\log_{10}(V_{\min}/\rho_{\phi0})$ close to its upper prior boundary,
which corresponds to a potential minimum close to the reference present-day
dark energy density. The correlations between the potential parameters and
the initial field displacement in Fig.~\ref{fig:galde_scalar_posterior_scatter}
illustrate how different scalar field configurations can reproduce similar
distance observables. Bounds adjacent to prior limits should therefore be
interpreted with their assumed prior ranges in mind.

\begin{table*}[ht!]
\centering
\scriptsize
\setlength{\tabcolsep}{5pt}

\begin{tabular}{llcccccc}
\hline
\hline
\addlinespace[2pt]
Dataset comparison & Model
& GoF & $\Delta{\rm GoF}$
& ${\rm S}$ & $\Delta{\rm S}$
& $-\ln{\rm R}$ & $\Delta(-\ln{\rm R})$ \\
\addlinespace[2pt]
\hline
\hline
\addlinespace[2pt]

\multirow{2}{*}{\shortstack[l]{CMB+BAO vs.\\PPS}}
& $\Lambda$CDM
& $5.472 \pm 0.038\,\sigma$
& --
& $5.289 \pm 0.037\,\sigma$
& --
& $27.83 \pm 0.76$
& -- \\

& Scalar field
& $5.5377 \pm 0.0055\,\sigma$
& $0.066 \pm 0.038\,\sigma$
& $5.3458 \pm 0.0028\,\sigma$
& $0.057 \pm 0.037\,\sigma$
& $27.21 \pm 0.93$
& $-0.6 \pm 1.2$ \\

\addlinespace[2pt]
\hline
\addlinespace[2pt]

\multirow{2}{*}{\shortstack[l]{CMB+BAO vs.\\CC+GRB+Pan}}
& $\Lambda$CDM
& $0.920 \pm 0.032\,\sigma$
& --
& $0.12 \pm 0.11\,\sigma$
& --
& $-5.4 \pm 1.7$
& -- \\

& Scalar field
& $0.9813 \pm 0.0084\,\sigma$
& $0.061 \pm 0.033\,\sigma$
& $0.075 \pm 0.041\,\sigma$
& $-0.05 \pm 0.12\,\sigma$
& $-7.3 \pm 3.5$
& $-1.9 \pm 3.9$ \\

\addlinespace[2pt]
\hline
\hline
\end{tabular}

\caption{\justifying{
Comparison of the goodness-of-fit (GoF),
suspiciousness statistic ${\rm S}$, and
Bayes ratio ${\rm R}$ for the two dataset concordance
analyses considered in this work.
The first comparison considers CMB+BAO versus
PPS, while the second considers
CMB+BAO versus CC+GRB+Pan.
Results are reported for $\Lambda$CDM and the
generalised axion-like dark energy model.
The quantities $\Delta{\rm GoF}$,
$\Delta{\rm S}$, and $\Delta(-\ln{\rm R})$
denote the differences between quintessence and $\Lambda$CDM.
Positive differences indicate increased discordance
relative to $\Lambda$CDM, whereas negative differences
indicate improved concordance.
}}
\label{tab:galde_dataset_tension}
\end{table*}

% --- Additional discussion: model comparison, dataset concordance, evolution ---
\subsection{Model comparison and dataset concordance}
\label{sec:galde_model_comparison}

We now assess whether the extra freedom of the scalar field is supported by
the data after accounting for model complexity. Throughout this comparison,
$\Delta X\equiv X_{\rm scalar}-X_{\Lambda{\rm CDM}}$, so that a
positive difference in the information criteria indicates a preference for
the simpler $\Lambda$CDM description. As shown in
Table~\ref{tab:galde_results}, $\Delta\mathrm{AIC}_{\rm C}$ is positive
for all six combinations, ranging from $5.749\pm0.011$ for PPS to
$15.690\pm0.097$ for BAO. The BIC differences are also positive throughout,
with particularly large penalties in the PPS and CMB+BAO+PPS fits,
$\Delta\mathrm{BIC}=22.031\pm0.011$ and
$22.398\pm0.028$, respectively. Thus, the improvement in the best-fit
likelihood is not sufficient to overcome the complexity penalty imposed by
these criteria.

The posterior-based criteria give a less pronounced, but broadly consistent,
picture. DIC differences lie between $0.38\pm0.17$ (PPS) and
$1.67\pm0.20$ (CMB+BAO+PPS), while WAIC differences are generally small;
in particular, BAO gives $\Delta\mathrm{WAIC}=-0.35\pm0.70$,
which is consistent with no preference. The values of
$\Delta(-\ln{\rm B})$ also depend on the dataset and have comparatively
large numerical uncertainties in several cases. For example, BAO gives
$-1.37\pm0.97$, whereas CMB+BAO gives $1.79\pm0.77$.
These evidence estimates should be interpreted cautiously because the
harmonic-mean evaluation is sensitive to low-likelihood posterior tails.
Taken together, the different criteria provide no robust statistical
preference for the additional scalar field degrees of freedom, even where
they broaden the range of allowed cosmological histories.

To investigate dataset consistency, we compare CMB+BAO separately with PPS
and CC+GRB+Pan, using the joint fits to evaluate the goodness-of-fit (GoF),
suspiciousness ($\rm S$), and Bayesian ratio ($\rm R$) in
Table~\ref{tab:galde_dataset_tension}. Both GoF and $\rm S$ are expressed in
equivalent Gaussian standard deviations. For CMB+BAO versus PPS, the
discrepancy remains strong in both models: the GoF significance changes
from $5.472\pm0.038\,\sigma$ in $\Lambda$CDM to
$5.5377\pm0.0055\,\sigma$ in the scalar field model, while the
corresponding $\rm S$ significance changes from $5.289\pm0.037\,\sigma$ to
$5.3458\pm0.0028\,\sigma$. The differences,
$\Delta\mathrm{GoF}=0.066\pm0.038\,\sigma$ and
$\Delta \rm S=0.057\pm0.037\,\sigma$, are small and do not point to any
alleviation of the Hubble tension. The Bayesian ratio difference,
$\Delta(-\ln \rm R)=-0.6\pm1.2$, is likewise compatible with zero.
The ability of PPS to accommodate a smaller $\Omega_{\rm m0}$ therefore
does not translate into a more concordant joint fit with CMB+BAO.

By contrast, CMB+BAO and CC+GRB+Pan are mutually consistent. The GoF
significance is $0.920\pm0.032\,\sigma$ in $\Lambda$CDM and
$0.9813\pm0.0084\,\sigma$ in the scalar field model, while the
corresponding suspiciousness significances are
$0.12\pm0.11\,\sigma$ and $0.075\pm0.041\,\sigma$.
Their differences, $\Delta\mathrm{GoF}=0.061\pm0.033\,\sigma$ and
$\Delta \rm S=-0.05\pm0.12\,\sigma$, are too small to establish a
meaningful change in concordance. The Bayesian ratio difference of
$-1.9\pm3.9$ is also inconclusive. These results are consistent with
the restricted scalar field parameter space inferred from the separate
CC+GRB+Pan fit: no substantial change to the background dynamics is
required for agreement with CMB+BAO.

These contrasting levels of dataset concordance are also illustrated
by the posterior distributions shown in Fig.~\ref{fig:galde_posterior_comparison},
where the tension between PPS and CMB+BAO is evident, while
CC+GRB+Pan exhibits substantial overlap with CMB+BAO.

\subsection{Reconstructed scalar field and density evolution}
\label{sec:galde_reconstructed_evolution}

\begin{figure*}[ht!]
    \centering

    % Top left: PPS equation of state
    \begin{subfigure}{0.49\textwidth}
        \centering
        \includegraphics[width=\linewidth]{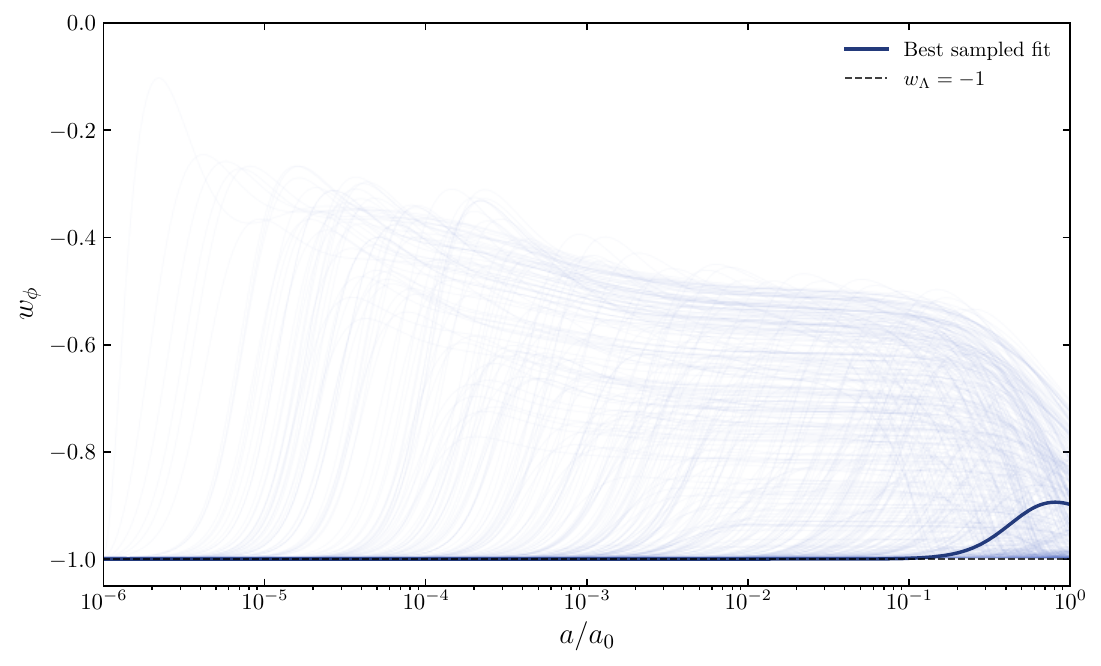}
        \caption{PPS: $w_\phi(a)$.}
        \label{fig:galde_wphi_pps}
    \end{subfigure}
    \hfill
    % Top right: CMB+BAO+PPS equation of state
    \begin{subfigure}{0.49\textwidth}
        \centering
        \includegraphics[width=\linewidth]{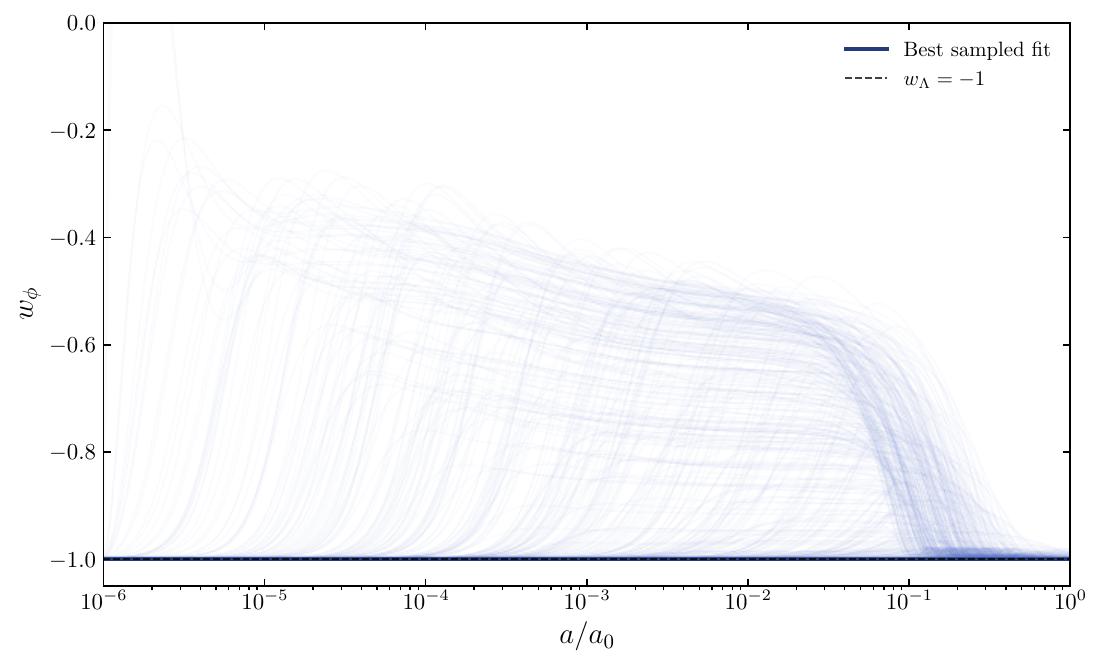}
        \caption{CMB+BAO+PPS: $w_\phi(a)$.}
        \label{fig:galde_wphi_cmb_bao_pps}
    \end{subfigure}

    \vspace{0.35cm}

    % Bottom left: PPS density parameters
    \begin{subfigure}{0.49\textwidth}
        \centering
        \includegraphics[width=\linewidth]{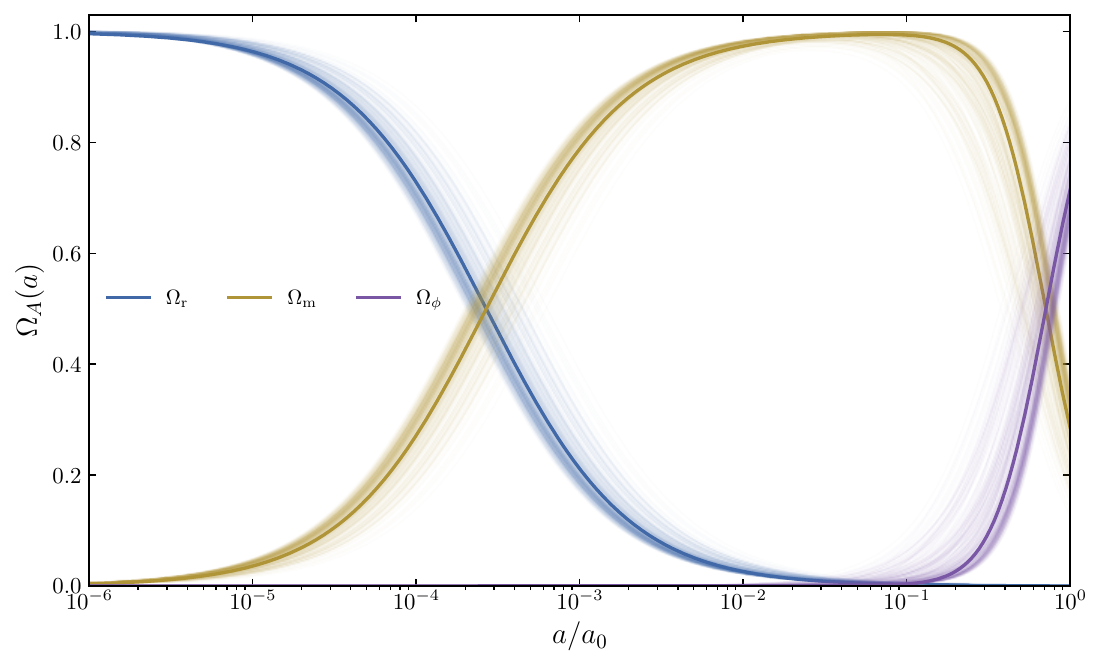}
        \caption{PPS: density parameters.}
        \label{fig:galde_omega_pps}
    \end{subfigure}
    \hfill
    % Bottom right: CMB+BAO+PPS density parameters
    \begin{subfigure}{0.49\textwidth}
        \centering
        \includegraphics[width=\linewidth]{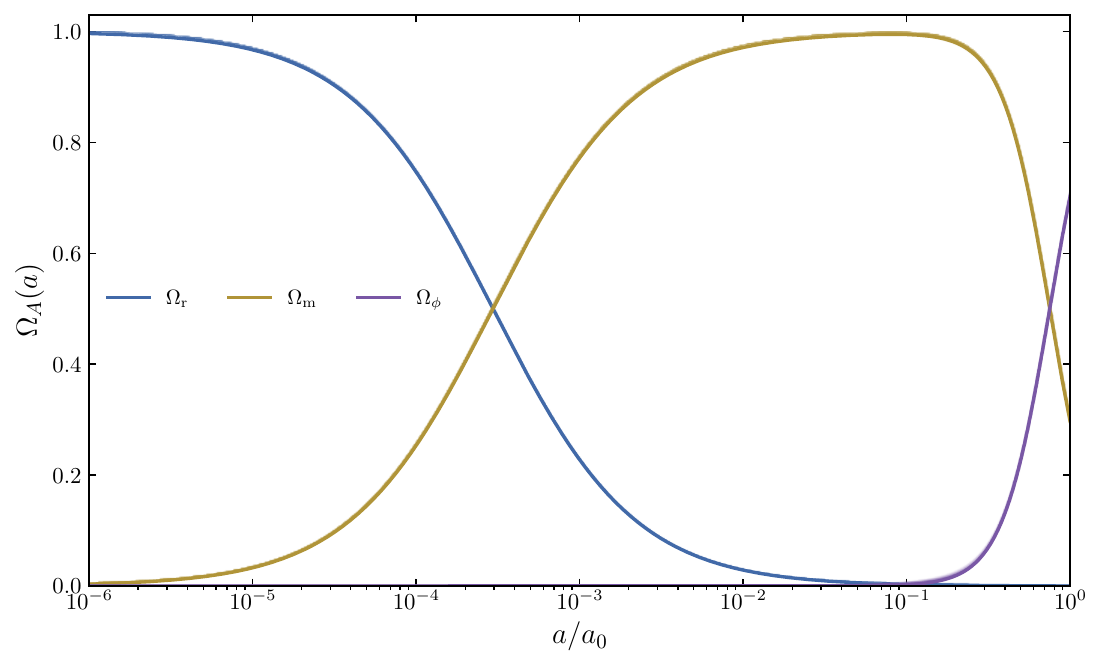}
        \caption{CMB+BAO+PPS: density parameters.}
        \label{fig:galde_omega_cmb_bao_pps}
    \end{subfigure}

    \caption{\justifying{
        Cosmological evolution of the generalised axion-like
        scalar field reconstructed from posterior samples
        obtained using PPS (left) and CMB+BAO+PPS (right).
        The upper panels show the evolution of the scalar field
        equation of state $w_\phi(a)$, with the blue curves
        representing posterior realisations and the darker
        solid curve corresponding to the best sampled fit.
        The dashed horizontal line indicates the cosmological
        constant value $w_\Lambda=-1$.
        The lower panels show the evolution of the fractional
        energy densities of radiation $\Omega_{\rm r}$ (blue),
        matter $\Omega_{\rm m}$ (gold), and the scalar field
        $\Omega_\phi$ (purple).
    }}
    \label{fig:galde_background_evolution_pps}
\end{figure*}

\begin{figure*}[ht!]
    \centering

    % Top left: CC+GRB+Pan equation of state
    \begin{subfigure}{0.49\textwidth}
        \centering
        \includegraphics[width=\linewidth]{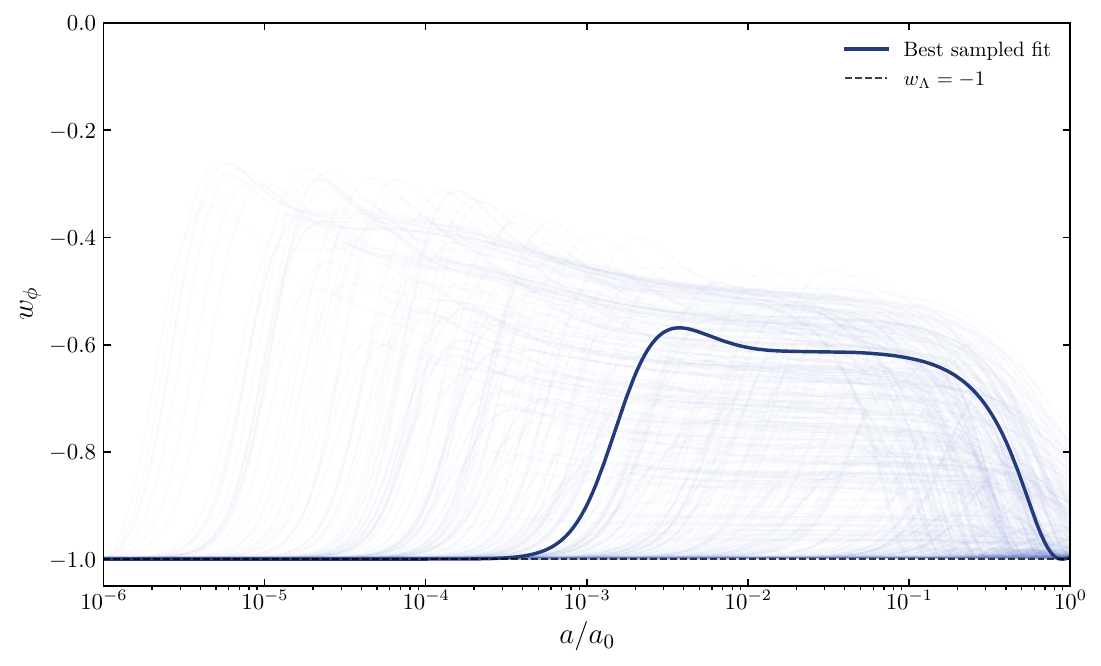}
        \caption{CC+GRB+Pan: $w_\phi(a)$.}
        \label{fig:galde_wphi_cc}
    \end{subfigure}
    \hfill
    % Top right: CMB+BAO+CC+GRB+Pan equation of state
    \begin{subfigure}{0.49\textwidth}
        \centering
        \includegraphics[width=\linewidth]{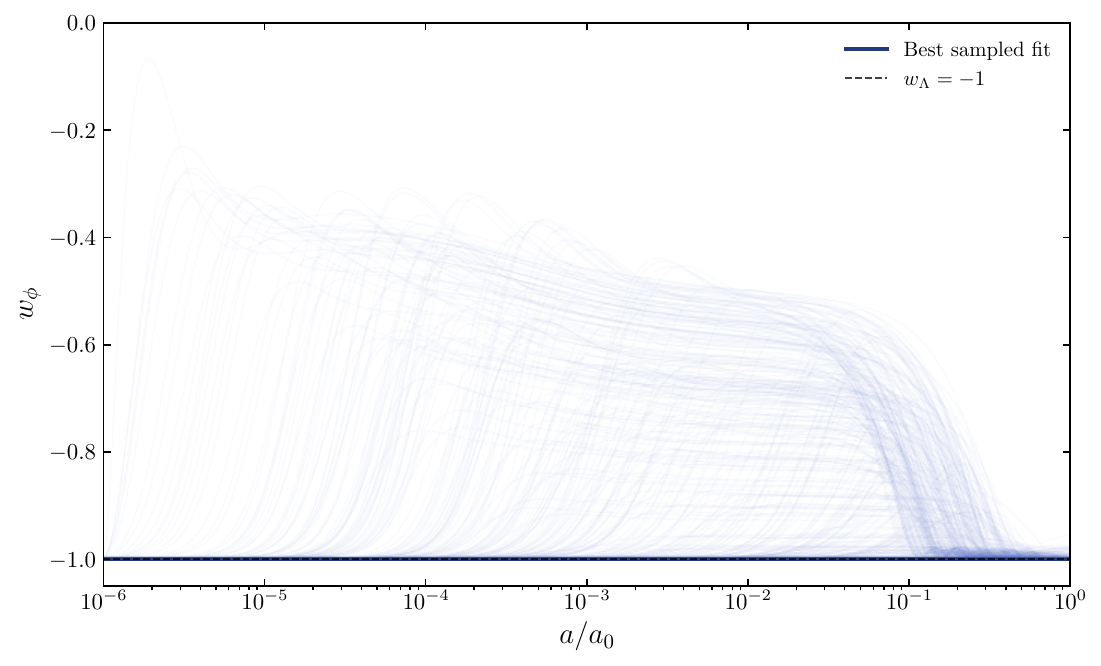}
        \caption{CMB+BAO+CC+GRB+Pan: $w_\phi(a)$.}
        \label{fig:galde_wphi_cmb_bao_cc}
    \end{subfigure}

    \vspace{0.35cm}

    % Bottom left: CC+GRB+Pan density parameters
    \begin{subfigure}{0.49\textwidth}
        \centering
        \includegraphics[width=\linewidth]{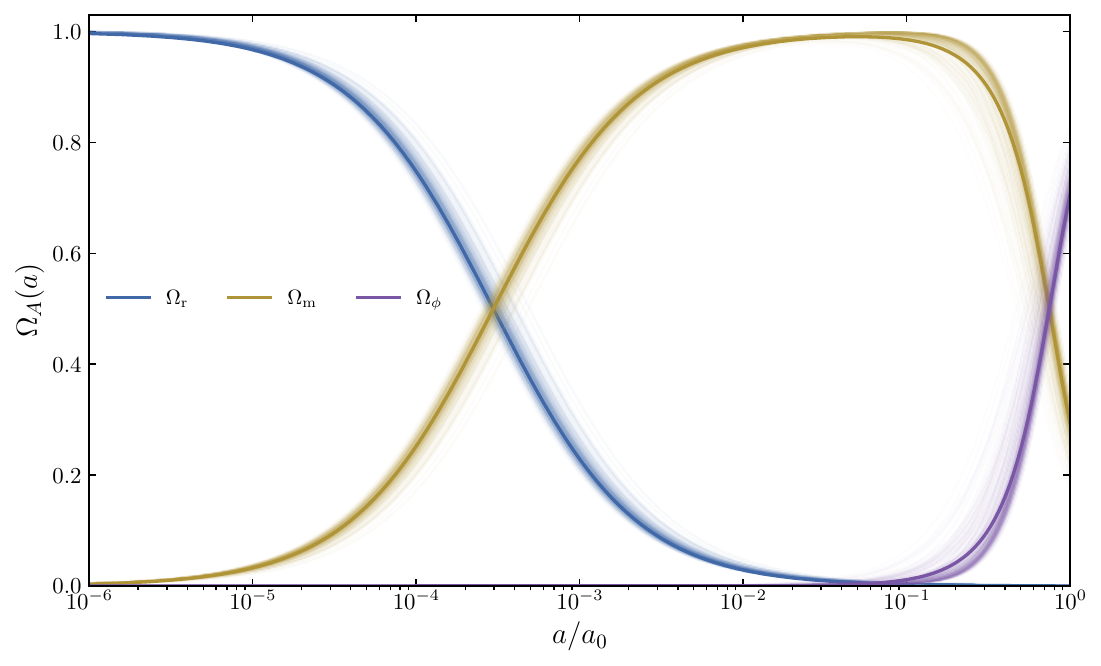}
        \caption{CC+GRB+Pan: density parameters.}
        \label{fig:galde_omega_cc}
    \end{subfigure}
    \hfill
    % Bottom right: CMB+BAO+CC+GRB+Pan density parameters
    \begin{subfigure}{0.49\textwidth}
        \centering
        \includegraphics[width=\linewidth]{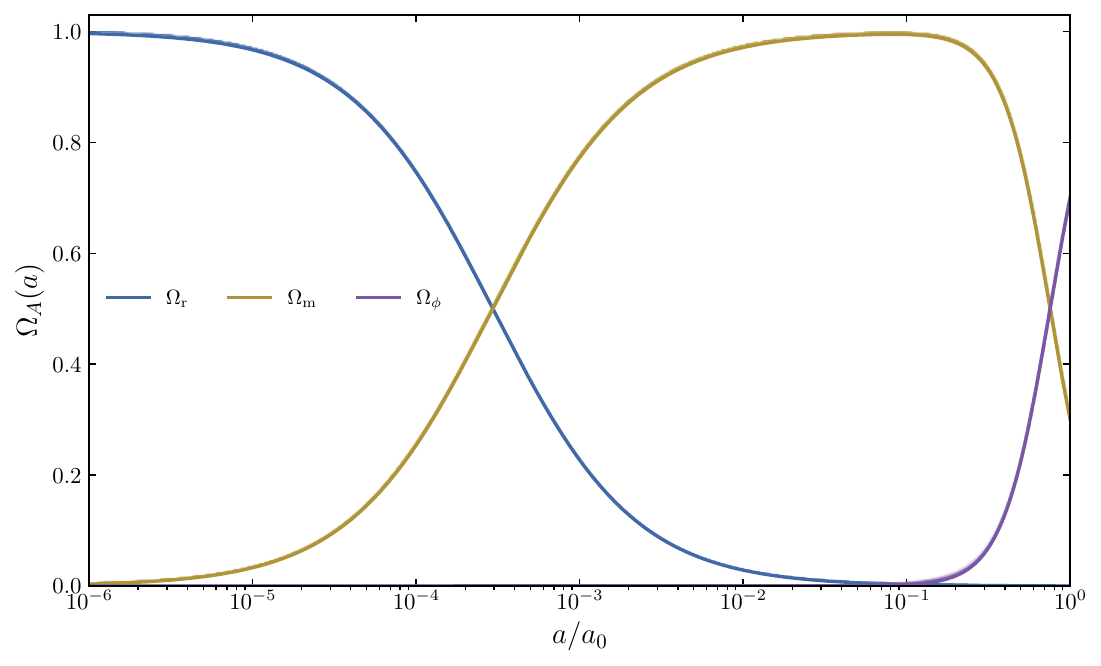}
        \caption{CMB+BAO+CC+GRB+Pan: density parameters.}
        \label{fig:galde_omega_cmb_bao_cc}
    \end{subfigure}

    \caption{\justifying{
        Cosmological evolution of the generalised axion-like
        scalar field reconstructed from posterior samples
        obtained using CC+GRB+Pan (left) and
        CMB+BAO+CC+GRB+Pan (right).
        The upper panels show the evolution of the scalar field
        equation of state $w_\phi(a)$, with the blue curves
        representing posterior realisations and the darker
        solid curve corresponding to the best sampled fit.
        The dashed horizontal line indicates the cosmological
        constant value $w_\Lambda=-1$.
        The lower panels show the evolution of the fractional
        energy densities of radiation $\Omega_{\rm r}$ (blue),
        matter $\Omega_{\rm m}$ (gold), and the scalar field
        $\Omega_\phi$ (purple).
    }}
    \label{fig:galde_background_evolution_cc}
\end{figure*}

Finally, we examine the background histories reconstructed from
posterior samples in Figs.~\ref{fig:galde_background_evolution_pps}
and~\ref{fig:galde_background_evolution_cc}. The upper panels show
the scalar field equation of state $w_\phi(a)$, while the lower
panels show the fractional densities $\Omega_{\rm r}(a)$,
$\Omega_{\rm m}(a)$, and $\Omega_\phi(a)$. These reconstructions allow
us to examine how the freedom in the scalar field dynamics
changes when the late-time datasets are combined with CMB+BAO.

We first consider PPS, shown in
Fig.~\ref{fig:galde_background_evolution_pps}. The PPS-only
constraints permit a broad variety of scalar field trajectories,
including substantial departures from $w_\phi=-1$. Such
behaviour is compatible with a temporary tracking phase,
whose approximate equation of state is
$w_\phi\simeq-(1+n)^{-1}$, followed by evolution towards
the potential minimum. The corresponding freedom in the
scalar field dynamics also allows a wider range of histories
for $\Omega_\phi(a)$. However, once CMB+BAO is included,
the allowed evolution becomes much more tightly restricted.
The posterior realisations cluster around histories close
to the $\Lambda$CDM limit, with only limited departures
from $w_\phi=-1$ remaining allowed. This illustrates that
the additional freedom permitted by PPS alone is strongly
constrained by the CMB+BAO information, consistent with
the persistence of the dataset tension discussed above.

A similar, although less pronounced, behaviour is found for
CC+GRB+Pan, as shown in
Fig.~\ref{fig:galde_background_evolution_cc}. This dataset
combination also permits temporary departures from
$w_\phi=-1$, but the range of allowed scalar field histories
is more restricted than in the PPS-only case. This is
consistent with the preference for a $\Lambda$CDM-like
branch already inferred from the corresponding parameter
constraints. Adding CMB+BAO further reduces the allowed
deviations, leading to a much narrower set of background
histories. In particular, the radiation-to-matter transition
and the onset of late-time scalar field domination become
very similar across the reconstructed posterior realisations.
The resulting evolution remains close to $\Lambda$CDM,
in agreement with the strong concordance between
CMB+BAO and CC+GRB+Pan.

These results illustrate that late-time observations alone
can accommodate non-trivial scalar field dynamics, with
the largest freedom found for PPS. Nevertheless, the
inclusion of CMB+BAO strongly restricts such deviations,
driving the allowed background evolution towards
$\Lambda$CDM-like histories. The same qualitative behaviour
is found when comparing BAO alone with CMB+BAO, although
we omit the corresponding reconstructions since they do
not provide additional information beyond the comparisons
shown here. The best sampled-fit trajectories should be
regarded as representative realisations rather than a
complete characterisation of the posterior uncertainty.
Overall, while temporary deviations from $w_\phi=-1$
remain possible, the statistical comparison does not
establish a preference for the additional scalar field
degrees of freedom.

\section{Conclusions}\label{conclusions}

Scalar fields have long played a central role in theoretical attempts to explain the late-time accelerated expansion of the Universe, offering a dynamical alternative to a rigid cosmological constant. Through an evolving equation of state, scalar field dark energy models can accommodate a wide range of expansion histories while addressing some of the conceptual challenges associated with $\Lambda$CDM. In this work, we have focused on a class of axion-like scalar field models characterised by a generalised potential that supports a prolonged tracking regime followed by late-time acceleration. By confronting this scenario with background cosmological observations from the cosmic microwave background (reproducing full Planck power spectrum result in \cite{Chiang:2025qxg}), baryon acoustic oscillations, Type~Ia supernovae, SH0ES determination of the local expansion rate, cosmic chronometers,
and gamma-ray bursts, we have assessed its phenomenological viability and shown that axion-inspired quintessence models provide a flexible and testable framework for describing the dark energy component driving the present cosmic acceleration. 

Our results show that the additional freedom introduced by the
scalar field allows departures from the $\Lambda$CDM expansion
history, particularly for the PPS (and to a lesser degree CC+GRB+Pan) dataset, which admits a
wider range of scalar field trajectories and lower values of
$\Omega_{\rm m0}$. However, these deviations become strongly
restricted when CMB+BAO information is included. Moreover,
the additional degrees of freedom do not alleviate the
discrepancy between CMB+BAO and PPS, which remains at
approximately $5.5\sigma$ in both models, nor disrupt the concordance between CMB+BAO and CC+GRB+Pan. The model
comparison statistics also provide no significant preference
for the generalised axion-like scenario over $\Lambda$CDM,
with the information criteria generally favouring the
simpler model. Therefore, although the scalar field can
exhibit tracking regime during matter-dominated era, current background observations
do not require departures from a cosmological constant. On the other hand, it does provide rich physics in earlier epoch that may serve as a platform for future extension, without any undesirable side effects.

\section*{Acknowledgements}

 M. B.-L. is supported by the Basque Foundation of Science Ikerbasque. HWC is supported by the NSFC Grant No.~12250410250, No.~12505074 and No.~12347133. C.G.B. acknowledges financial support from the French National Research Agency (ANR) under project No. ANR-25-CE57-1445-01 and from the Spanish Ministry of Science, Innovation and Universities under the FPI fellowship PRE2021-100340. T.J.B.  is supported by the Basque Foundation of Science Ikerbasque. Our work is supported by the Spanish Grant PID2023-149016NB-I00 (MINECO/AEI/FEDER, UE). This work is also supported by the Basque government Grant No. IT1628-22 (Spain). The authors acknowledge the contribution of the COST Action CA21136 “Addressing observational tensions in cosmology with systematics and fundamental physics (CosmoVerse)”.
\bibliography{merged_bibliography}

\end{document}